\documentclass{aa}  

\usepackage{graphicx}
\usepackage{txfonts}
\usepackage{lipsum}
\usepackage{subcaption}         

\usepackage{lscape}
\usepackage{placeins}
                                
\usepackage[colorlinks=true,
   linkcolor=blue, citecolor=blue, filecolor=blue, urlcolor=blue]{hyperref}

\begin{document}

\title{A dedicated survey of fast-rotating near-Earth asteroids with the Two-meter Twin Telescope}
\subtitle{I. Observational strategy and first results}

   \author{M.R. Alarcon\inst{1,2,3}\fnmsep\thanks{Corresponding author: mra@mralarcon.com}
        \and J. Licandro\inst{1,2}
        \and M. Serra-Ricart\inst{1,2,3}
        }
   \authorrunning{Alarcon et al.}
   \titlerunning{A dedicated survey of fast-rotating NEAs} 
   
   \institute{Instituto de Astrofísica de Canarias (IAC), C/ Vía Láctea, s/n, E-38205, La Laguna, Tenerife, Spain
            \and Departamento de Astrofísica, Universidad de La Laguna (ULL), E-38206 La Laguna, Tenerife, Spain
            \and Light Bridges, Observatorio Astronómico del Teide. Carretera del Observatorio del Teide, s/n, E-38570 Güímar, Tenerife, Spain
            }

   \date{Received 11 December 2025 / Accepted 5 February 2026}

\abstract
 {The rotational properties of small near-Earth asteroids (NEAs) provide crucial insights into their internal structure and collisional history. However, systematic surveys targeting metre- to decametre-sized bodies are rare, thus leaving their spin distribution poorly constrained.}
 {Our aim was to quantify the prevalence of fast rotation and characterise the spin-rate distribution of small NEAs to constrain their internal strength and evolution.}
 {We conducted a dedicated high-cadence photometric survey of 249 NEAs using the Two-meter Twin Telescope (TTT). Rotation periods and amplitudes were derived from dense time series to classify objects as fast or non-fast rotators.}
 {We determined the rotation period of 156 new fast rotators (P < 2.2 h), including 87 that rotate faster than 10 min. The prevalence of fast rotators increases with absolute magnitude: from 60.6--80.3 \% for 22 < H < 24 to 77.3--89.4 \% for 24 < H < 26 and 94.1--96.1 \% for H > 26, indicating that fast rotation dominates in the small NEA population. Most objects spin faster than the gravity-defined limit; 98 targets require cohesive strengths exceeding that of weak rubble piles, and 22 are compatible only with compact, high-strength interiors.}
 {This is the first systematic survey targeting the rotation of such small NEAs, providing the largest homogeneous sample of fast rotators obtained by a single campaign. Our findings demonstrate that fast rotation is the norm for objects smaller than tens of metres, implying that modest cohesive strength is required to prevent their rotational disruption.}

   \maketitle
\nolinenumbers

\section{Introduction}
The rotational properties of asteroids offer crucial insights into their internal structure, formation history, and evolutionary processes. For an asteroid conceived as a homogeneous rubble pile, a loose aggregate of rocks held together primarily by self-gravity, there exists a critical spin rate at which the centrifugal forces at the equator overcome the inward pull of gravity \citep{Pravec2000}. Beyond this rate, the object is unstable, and undergoes mass shedding, reshaping, or eventual breakup \citep{Walsh2008, Harris2009, Yu2018, Zhang2018, Zhang2021, Cheng2021}. This critical limit is known as the cohesionless spin barrier, which, for typical asteroid bulk densities, corresponds to a rotation period of approximately 2.2 hours. The absence of asteroids larger than a few hundred metres in diameter rotating faster than this threshold--with a few exceptions (see \cite{Chang2017,Chang2019,Chang2022,Strauss2024})--has long been regarded as evidence that most asteroids are gravity-dominated rubble piles.

The situation changes dramatically for smaller bodies. Surveys have revealed that near-Earth asteroids (NEAs) smaller than $\sim$200~m often rotate well above the 2.2 h barrier \citep{Whiteley2002, Warner2009, Statler2013, zambrano2024}. These fast-rotating asteroids (FRAs) demonstrate that another source of strength must be at play. Laboratory experiments, numerical models, and meteoritic analyses indicate that weak cohesive forces, particularly van der Waals interactions between fine grains, can provide tensile strengths of tens to hundreds of pascals, sufficient to stabilise a 100 m body with a rotation period of just a few minutes \citep{Holsapple2007, Scheeres2010, Sanchez2014}. This implies that cohesion, apart from gravity, governs the structural stability of the smallest asteroids. 

While previous efforts have provided valuable insights into this population, a comprehensive survey focused on determining the prevalence of fast-rotating objects has remained a challenge. Early work by \cite{Hergenrother2011} surveyed 53 small NEAs, and found that a minimum of two-thirds of objects with absolute magnitude H > 20 are fast rotators with periods significantly faster than 2~h. They found that this percentage increases with decreasing size and that a minimum of 79\% of H $\geq$ 24 objects are fast rotators. Subsequent characterisation campaigns, such as the MANOS survey \citep{Thirouin2016, Thirouin2018} or the study by \cite{Sanchez2024}, provided valuable data on a wider range of physical properties, including composition and rotational light curves. However, these surveys were not specifically designed to assess the prevalence of fast rotators and some (e.g. MANOS), while targeting similarly small fast-moving asteroids with trailing-limited exposures, may have had their sensitivity to the shortest-period variations constrained by effective cadence and readout overheads. In addition to broad surveys, numerous dedicated studies have focused on characterising individual objects of particular interest \citep{Licandro2023,Popescu2023,Beniyama2024}. More recently, specialised observational techniques have been employed to find the fastest rotators. For example, the Tomo-e Gozen video observations by \cite{Beniyama2022} confirmed the existence of a population of tiny fast rotators that had been missed by previous surveys, and the approach of trailed photometry by \cite{Devogele2024} demonstrated a new way to perform accurate photometry of fast-moving objects, leading to the detection of asteroid 2024 BX1 with a rotation period of just $2.5888\pm0.0002~$s, the fastest ever recorded. 

A significant fraction of our current knowledge stems from long-term efforts to compile and maintain rotational data, most notably the Light Curve Database (LCDB, \citealt{Warner2009}) and the SsODNet database \citep{Berthier2023}, which aggregates thousands of published light curves into a single accessible resource. A wealth of additional information exists in the form of disseminated results and unpublished observations, for instance the extraordinary collection of light curves obtained over decades by Petr Pravec\footnote{\url{https://space.asu.cas.cz/~ppravec/}} and collaborators. These collective efforts provide the essential baseline for understanding the rotational behaviour of NEAs. Nevertheless, the prevalence of fast rotators among the small NEA population has not been explicitly stated in previous works, as no survey has been conducted with this specific aim.

In this work we present a dedicated survey of small near-Earth asteroids aimed at establishing the prevalence of fast rotators within this population. We describe in Sect.~\ref{sec:methods} the observing strategy and analysis methods adopted to efficiently detect rapid rotation among faint fast-moving targets. The observational results are presented in Sect.~\ref{sec:results}, and in Sect.~\ref{sec:prev} we quantify the prevalence of fast rotation as a function of size. Finally, in Sect.~\ref{sec:discussion} we analyse the observational biases affecting our survey and place these findings in the context of previous work and of the physical properties and evolutionary pathways of small NEAs.

\section{Methods} \label{sec:methods}
\subsection{Two-meter Twin Telescope}
This survey is based on observations carried out since late 2023 with the Two-meter Twin Telescope (TTT\footnote{\url{https://ttt.iac.es/}}), located at the Teide Observatory, Canary Islands, Spain (28$^\circ$18$'$04$''$ N, 16$^\circ$30$'$38$''$ W, alt. 2362 m). The facility consists of two identical 2.0 m f/6 Ritchey-Chrétien telescopes (TTT3/4) and two smaller replicas of 0.80 m f/6.85 (TTT1/2), all mounted on Alt-Az systems and equipped with multiple instruments at the Nasmyth foci. The 0.80 m telescopes have been operational since 2023 (MPC codes Y65, Y66), while the first 2.0 m unit (TTT3, Y68) has been under commissioning since early 2025.  All telescopes are equipped with the Fast Embedded-sCMOS Robotic Visible Observatory for Rapid transients (FERVOR), an optical imager based on sCMOS sensors (Sony IMX455 and IMX411; \citealt{Alarcon2023}). The large-format version, installed on TTT1/2, provides a field of view of $33.3'\times25.0'$ with a plate scale of $0.14''~{\rm px^{-1}}$, while the medium-format version, installed on TTT3, provides $10.2'\times6.8'$ with a plate scale of $0.19''~{\rm px^{-1}}$ when operated with $3\times3$ binning. These detectors have negligible dark current, virtually no readout overhead, a minimum exposure time of 1 s, and very low readout noise ($1.8~e^-$ for the large version and $4.7~e^-$ for the medium version). To maximise the signal-to-noise ratio (S/N), observations were performed with a broadband UV-IR Cut filter \citep{ATLASTeide}, with a flat effective transmission in the range 400--700 nm, approximately equivalent to the Sloan $g'+r'$ bands.

For a small subset of targets, we also made use of commissioning time of the Transient Survey Telescope (TST,\footnote{\url{https://tst.iac.es}} Y64), a 1.0 m f/1.3 wide-field telescope located near TTT at the same observatory. TST is equipped with the same large-format FERVOR-L camera as TTT1/2, which provides a field of view of $2.3^\circ\times1.7^\circ$ and a plate scale of $0.598''~{\rm px^{-1}}$.

\subsection{Survey strategy}
\begin{figure}[t!]
\centering
\includegraphics[width=\hsize]{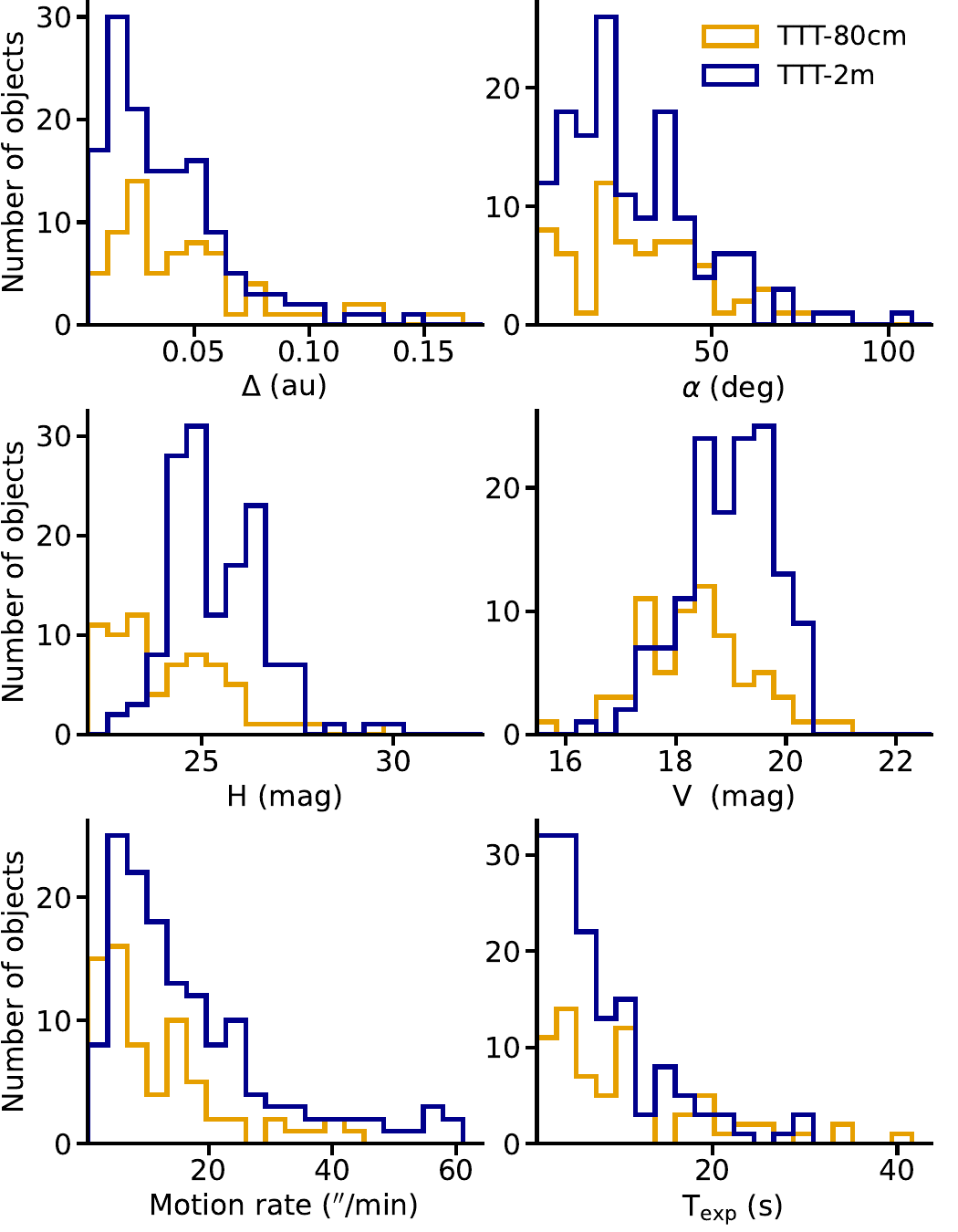}
  \caption{Distributions of observational parameters for the targets in our survey. The panels show (from left to right and top to bottom) geocentric distance $\Delta$, phase angle $\alpha$, absolute magnitude H, apparent V magnitude, sky-plane motion rate, and exposure time. Data obtained with the 0.8 m telescopes (TTT1/2, orange) and the 2 m TTT3 (dark blue) are shown separately.}
     \label{fig:ORs}
\end{figure}

A list of observable NEAs from the Teide Observatory was generated every day using the JPL Horizons system \citep{Giorgini1996}. The TTT facility provides an API-accessible Exposure Time Calculator (ETC), which is used to verify that the expected S/N exceeds 20 for at least 30 minutes of observation. The ETC accounts for the telescope’s mechanical limits, the target’s apparent magnitude and sky-plane angular rate, as well as the predicted sky brightness and seeing conditions. All candidates with absolute magnitude H > 22 are preselected for the 0.80 m telescopes (TTT1/2). For these objects, the S/N is evaluated and, if it is high enough, an observing run (OR) of two hours total duration is generated. Targets with H > 24 that cannot be observed with the smaller telescopes are assigned to the 2 m unit (TTT3), for which ORs of 40 minutes are created if the predicted S/N is above the threshold. In all cases, the selected exposure time is the maximum allowed by the object's motion and the night seeing. This process is carried out automatically every day. The final scheduling of ORs throughout the night is managed by the telescope operations team. Figure~\ref{fig:ORs} provides an overview of the survey sample, displaying the distributions of apparent and absolute magnitudes, phase angles, geocentric distances, angular rates, and exposure times of the observed targets.

\subsection{Data processing}
All raw frames were corrected for bias and twilight flat-field variations using standard procedures. Point-source photometry was then performed with a GPU-accelerated custom pipeline that employs PSF-matched convolutions for source detection and adaptive, S/N-driven aperture photometry with PSF-based aperture corrections. For each exposure, field stars were matched to the Pan-STARRS DR1 (PS1) catalogue \citep{Chambers2016} to derive a photometric zero-point, using a linear combination of the $g$ and $r$ bands: 0.516$\cdot$PS1$_{\rm g}$ + 0.484$\cdot$PS1$_{\rm r}$. The photometry of all detected point-sources is stored in a SQL-like database. This constitutes the standard pipeline for TTT images and runs in real time, immediately after the end of each exposure. The pipeline was validated in previous works \citep{Popescu2025,3I2025} and will be released as open-source software next year (Alarcon et al., Lemes-Perera et al., in prep.).

At the end of each night, the predicted positions of the targets from the JPL Horizons system were cross-matched with the photometric database. The photometric data were visually inspected, and outliers caused by asymmetric PSFs, streaks, or blending with background sources were rejected. Only measurements with photometric uncertainties smaller than 0.1 mag were retained. The time stamps of all detections were corrected to mid-exposure and for light-travel time.

To combine data from different nights, the magnitudes were normalised for heliocentric and geocentric distances and corrected for phase angle, following the H--G system of \citet{Bowell1989}. We applied three standard corrections:

First, a distance correction,
\begin{equation}
\delta V_{\rm dist} = -5 \log_{10}(r \cdot \Delta),
\end{equation}
where $r$ and $\Delta$ are the heliocentric and geocentric distances (in au). Second, a phase-angle ($\alpha$) correction,
\begin{equation}
\delta V_{\alpha} = 2.5 \log_{10}\left[(1-G)~\phi_1(\alpha) + G~\phi_2(\alpha)\right],
\end{equation}
where the basis functions are given by
\begin{equation}
\phi_i(\alpha) = \exp\left\{-A_i \left[\tan\left(\frac{\alpha}{2}\right)\right]^{B_i}\right\},
\end{equation}
with $(A_1,B_1)=(3.33,0.63)$ and $(A_2,B_2)=(1.87,1.22)$ and $G=0.15$. Additionally, we corrected the mid-exposure Julian Date for light-travel effects using
\begin{equation}
\delta t = -0.005778 \cdot \Delta.
\end{equation}

Rotational periods were derived through a two-step procedure. First, a fourth-order Lomb-Scargle periodogram \citep{Lomb1976,Scargle1982} was computed to identify candidate frequencies. This method generalises the classical Fourier periodogram to irregularly sampled data by fitting sinusoids at each trial frequency and evaluating the associated reduction in $\chi^2$. The frequency of the highest-power peak was adopted as the initial estimate. In the second step, the period was refined using phase dispersion minimisation (PDM; \citealt{Stellingwerf1978}), which evaluates the variance of phase-folded data across trial periods. The best period corresponds to the minimum of the PDM statistic, and its uncertainty was estimated from the half-width at half-minimum of the function around the solution. A third-order Fourier series was then fitted to the folded light curve to provide a smooth model.

The peak-to-peak amplitude of the Fourier model was adopted as the measured light curve amplitude. Because the exposure time T$_{\rm exp}$ is non-negligible compared to the rotation period P$_{\rm rot}$ for some of our targets, the observed variations are averaged over the integration, leading to a systematic underestimation of the true amplitude. This   amplitude smearing acts as a low-pass filter on the light curve and becomes increasingly relevant for faster rotators. To correct for this effect, we applied the standard sinc-function correction factor derived from integrating a sinusoidal signal over the exposure time,
\begin{equation}
A = A_{\rm obs}~\frac{\pi T_{\rm exp}/P_{\rm rot}}{\sin\left(\pi T_{\rm exp}/P_{\rm rot}\right)} ,
\end{equation}
where A$_{\rm obs}$ is the observed peak-to-peak amplitude. For P$_{\rm rot} \gg$ T$_{\rm exp}$ the correction factor approaches unity and the smearing effect is practically unnoticeable, whereas for P$_{\rm rot}$ of the order of T$_{\rm exp}$ the correction becomes significant.

Finally, assuming that the asteroid is a triaxial ellipsoid with semi-axes $a \geq b \geq c$, a lower limit on the axis ratio can be estimated from the corrected amplitude following
\begin{equation}
\frac{a}{b} \geq 10^{0.4A}.
\end{equation}

\section{Results} \label{sec:results}

\begin{figure*}[t]
\centering
\begin{minipage}{0.49\textwidth}
    \centering
    \includegraphics[width=\textwidth]{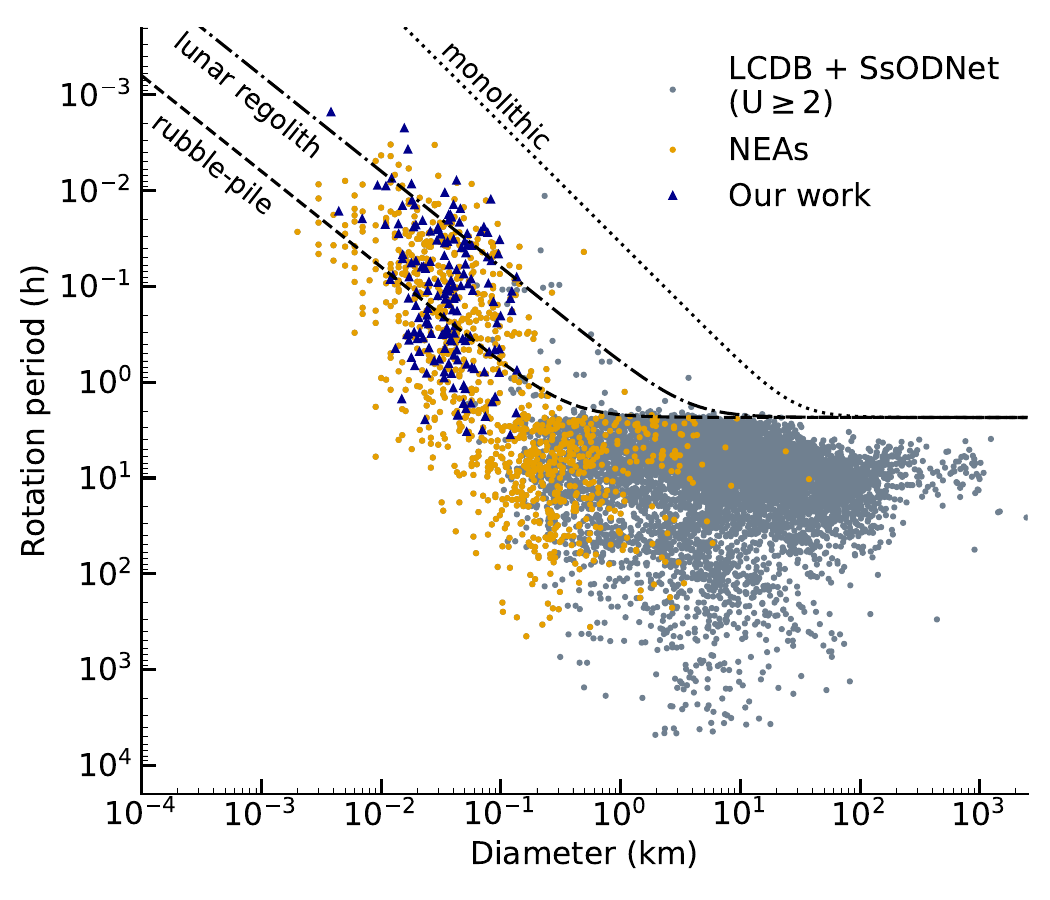}
\end{minipage}
\hfill
\begin{minipage}{0.49\textwidth}
    \centering
    \includegraphics[width=\textwidth]{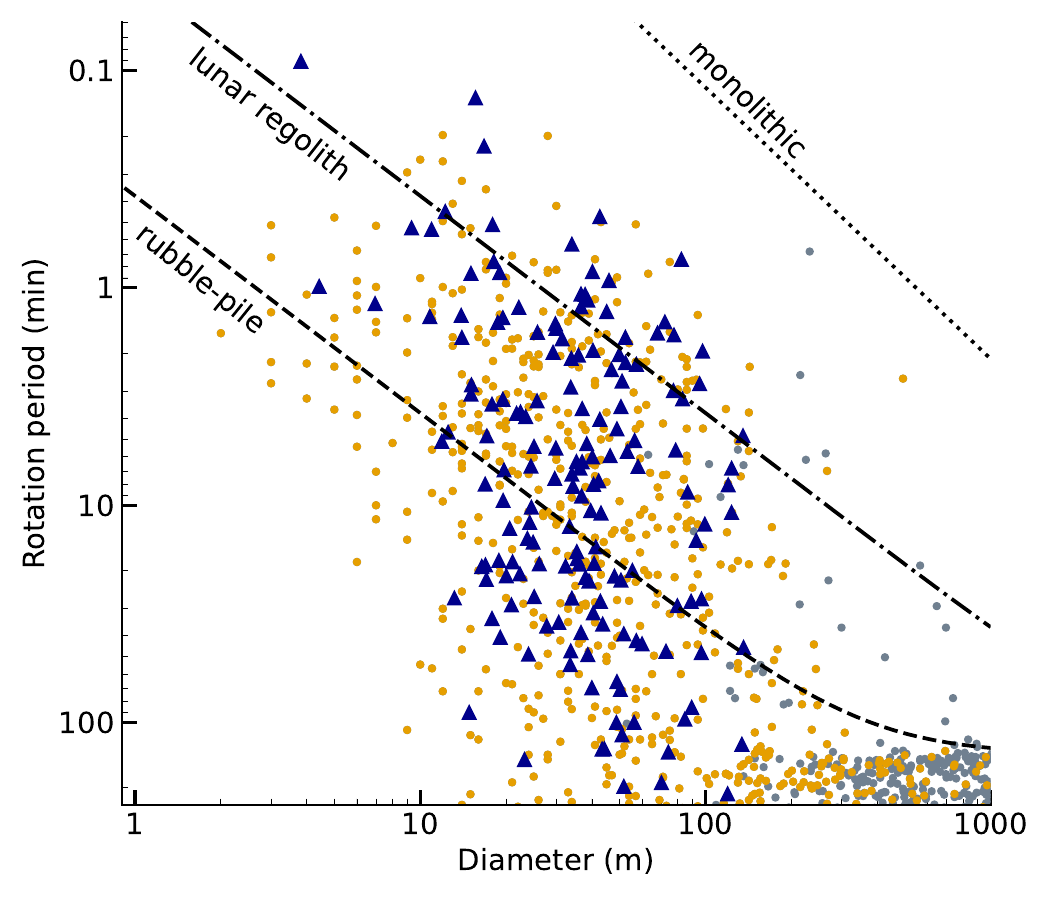}
\end{minipage}
\caption{
Rotation period as a function of diameter for the objects in this work. The left panel shows the full distribution, with asteroids from the Lightcurve Database (LCDB) and SsODNet with $U \geq 2$ in grey and near-Earth asteroids (NEAs) in orange. The 161 new NEAs with rotation periods derived in this paper are plotted in dark blue. Diameters were estimated from absolute magnitudes assuming a geometric albedo $p_V = 0.15$. Cohesive spin limits for weak rubble-pile material ($C\sim10$~Pa), lunar-like regolith ($C\sim1$~kPa), and a size-dependent monolithic strength ($C \propto D^{-0.5}$) are overplotted for reference. The right panel provides a zoomed-in image of the sub-140~m size range, where our survey is primarily focused.
}
\label{fig:spin_diagram}
\end{figure*}

We observed a total of 249 near-Earth asteroids during the survey. For 161 of them, we derived reliable rotation periods, of which 156 exhibit periods shorter than 2.2 h and are therefore considered fast rotators, while only five objects rotate more slowly. Figure~\ref{fig:spin_diagram} presents the distribution of rotation period as a function of diameter for our sample. The objects populate the fast-rotator regime almost homogeneously across the entire period range, from the classical spin barrier down to periods of a few seconds. The observational and derived parameters of most of these objects are listed in Table~\ref{tab:periods}, and their corresponding phased-folded light curves are presented in Appendix~\ref{sec:append_LC}. Diameters were estimated from the absolute magnitude using
\begin{equation}
D = \frac{1336}{\sqrt{p_V}}~10^{-0.2H},
\end{equation}
assuming a geometric albedo $p_V = 0.15$, a commonly adopted representative value for small NEAs.

To place these results in context, we overplot the NEAs listed in the LCDB and SsODNet with period quality $U\geq2$, comprising 520 fast rotators. Our survey increases the number of well-characterised fast-rotating NEAs by 30\%, particularly in the sub-100~m regime. In particular, 87 objects show super-fast rotation, with periods shorter than 10~min, placing them among the most rapidly rotating NEAs known. This represents a substantial expansion of the known population in this extreme rotation range.\\

Such extremely short rotation periods raise important questions about the internal structure and strength of these bodies, as gravitational binding alone cannot provide sufficient stability at these spin rates. To investigate this, we include in Fig.~\ref{fig:spin_diagram} the cohesion curves computed following the spin limit for cohesive bodies derived from the Drucker--Prager yield criterion \citep{Holsapple2007}. In this framework, the internal strength arises from friction and cohesion, allowing the material to sustain stresses that exceed those of a purely gravity-dominated body. Experimental and theoretical analyses indicate that these contacts can provide tensile strengths of order $C\sim10$~Pa for rubble aggregates covered by micron-sized grains \citep{Scheeres2010} and up to $C\sim1$~kPa for regolith with lunar-like packing properties \citep{Sanchez2014}. For fractured monolithic material, impact experiments suggest a strength scaling $C\propto D^{-0.5}$ \citep{Housen1999}, implying substantially higher effective cohesion at tens of metres in size and allowing rotation periods of a few tens of seconds.

The limiting spin period for a cohesive spherical body can be written, following the limit-analysis formulation of \citet{Holsapple2007} and as presented in its closed form by \citet{Zhang2021}, as
\begin{equation}
P_{\mathrm{rot, limit}} =\pi R ~\sqrt{
  \frac{6\rho(\sin\phi + 1)}
       {4\pi G \rho^{2} R^{2} \sin\phi + 15 C\cos\phi}} ~,
\label{eq:Plimit_cohesion}
\end{equation}
where $R$ is the radius, $\rho$ the bulk density, $\phi$ the internal friction angle, and $C$ the cohesive strength. This expression highlights the size dependence of the strength-controlled regime: as $R$ decreases, the cohesive term in the denominator becomes dominant, allowing substantially shorter limiting periods even for modest values of $C$. For the curves shown in Fig.~\ref{fig:spin_diagram}, we adopt $\rho=3~\mathrm{g~cm^{-3}}$ and $\phi=30^\circ$, representative of ordinary chondritic material and typical regolith friction angles. 
These theoretical curves represent upper bounds to the actual spin limit, as they rely on volume-averaged stress conditions. Numerical analyses show that yielding typically initiates locally before the global failure criterion is reached, meaning that real bodies may disrupt at slower spins \citep{Zhang2021}.

In this context, 98 of our targets lie above the limit expected for a weak rubble pile, requiring at least regolith-level cohesion or fractured monolithic structure; 22 of them exceed the monolithic strength curve and are only compatible with a compact, high-strength interior. The largest object in this group is 2024 QS1, with H = 22 and an estimated diameter of $\sim140$~m, rotating in only $4.803\pm0.034$~min. Its location well above the rubble-pile and regolith-strength curves strongly suggests a monolithic nature or, at least, a highly cohesive internal structure. Our fastest detected rotator is 2025 WR7 (H = 29.8), with a period of $5.44\pm0.22$~s.\\

Among the 156 new fast rotators, we identify at least 35 objects displaying non-principal axis rotation (tumbling). This number is unexpectedly large when compared to the LCDB, which lists only 12 tumblers among its 505 fast rotators. Our survey potentially increases the known population of fast-rotating tumblers by almost a factor of three, raising important questions about the true fraction of fast rotators that may actually be tumblers.

A more detailed analysis of these objects, including objects identified in previous surveys but not flagged as such in the LCDB (e.g. \citealt{Thirouin2018, Sanchez2024}), will be presented in a separate paper. That study will investigate the interplay between spin acceleration and the damping of non-principal axis rotation in the fast-rotation regime, a combination that has not yet been explored in detail for small NEAs.

For the prevalence estimates discussed below, these objects are counted as fast rotators. However, their individual properties and light curves are not included in Table~\ref{tab:periods} and Fig.~\ref{fig:LC-1}.

\section{Prevalence}\label{sec:prev}
To estimate the prevalence of fast rotators as a function of size, we considered not only the objects with confirmed rotation periods (including tumblers), but also those for which a reliable period could not be derived. Although these targets do not meet the criteria for a secure period determination, their light curves still provide constraints on whether they are likely fast rotators or not, and they can therefore be included in the statistical analysis.

\begin{figure}[t!]
\centering
\includegraphics[width=\hsize]{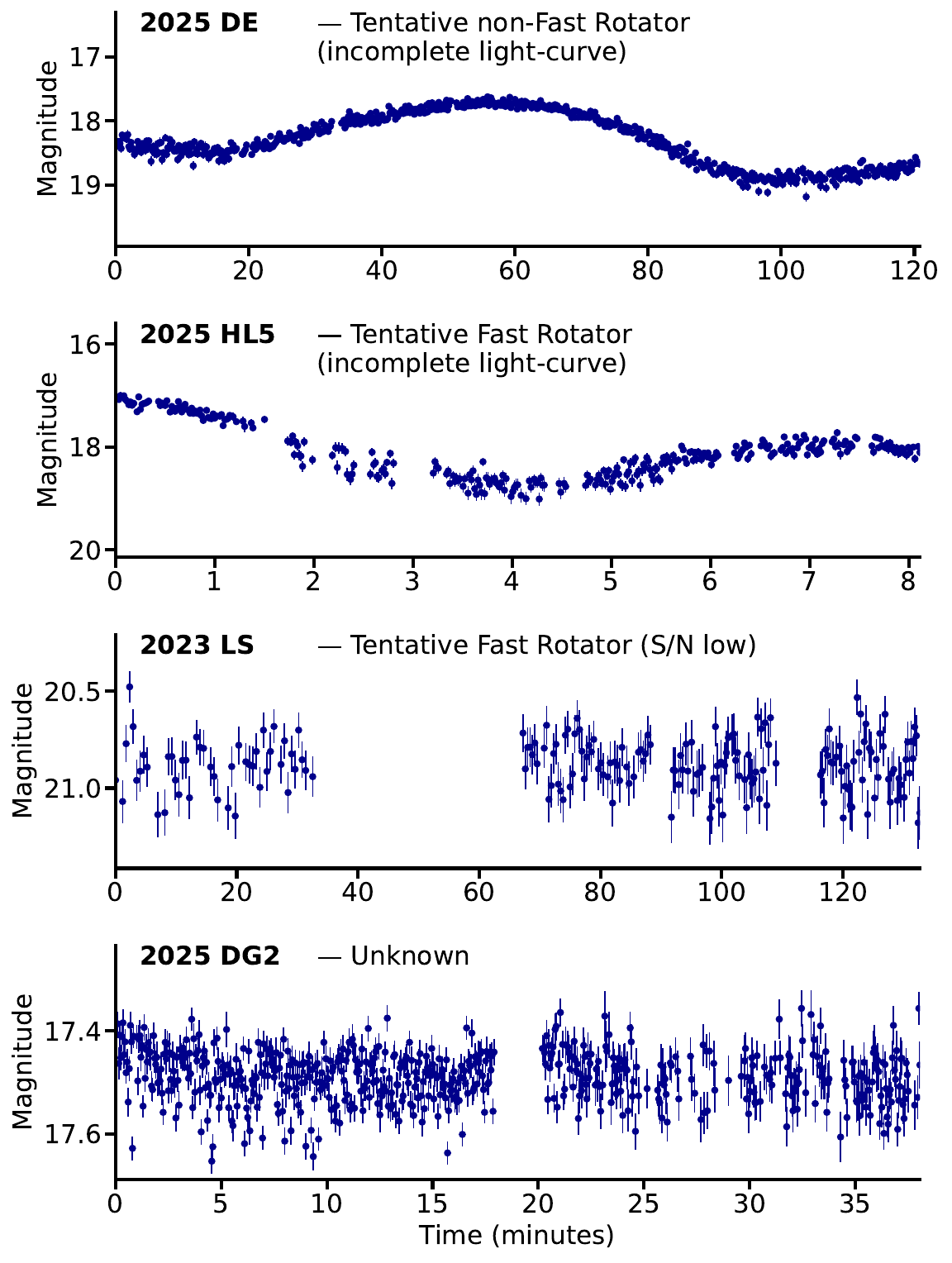}
  \caption{Representative examples of light curves without confirmed rotation periods, illustrating the types of partial or S/N-limited data included in the prevalence analysis.}
     \label{fig:not_confirmed}
\end{figure}

Figure~\ref{fig:not_confirmed} illustrates four representative cases drawn from this group.  
The first two examples correspond to incomplete light curves, typically resulting from observing runs shorter than the object's rotation period, either by survey design (40--120 min) or telescope time availability. The object 2025 DE (H=23.9) shows an incomplete light curve, yet its slow and monotonic trend makes it straightforward to rule out a fast-rotating nature (the partial curve is consistent with a period of $\sim$150~min). Conversely, 2025 HL5 (H=23.7) exhibits clear variability within the short observing window, making it easy to identify it as a likely fast rotator (with a period of approximately 12~min).
A different situation is illustrated by 2023 LS (H=29.9), whose light curve shows variations and a large robust amplitude ($\Delta m_{\rm P95-P5}\simeq 0.45$~mag). In segments of the data with higher S/N, quasi-periodic trends are apparent; however, no single rotation period provides a coherent phase-folded solution across the full observing run, and the best-fitting periods remain non-unique. We therefore consider 2023~LS a tentative fast-rotator candidate, as its high-amplitude modulation is clear, although the full dataset cannot be coherently phase-folded with a single period. In contrast, 2025 DG2 (H=24.9) displays only low-amplitude variability, with a small robust amplitude ($\Delta m_{\rm P95-P5}\simeq 0.16$~mag) and no repeatable structure emerging under phase-folding. In such low-amplitude cases, both long rotation periods and insufficient S/N remain plausible, and even smearing effects cannot be excluded; we therefore classify 2025~DG2 as unknown.\\

To quantify how common fast rotation is as a function of size, we considered both the confirmed fast rotators and the objects classified as tentative fast rotators. A subset of the sample remains unclassified due to the absence of detectable variability or insufficient photometric precision. Because these unknown objects may include both fast and non-fast rotators, we report an interval for the fast-rotator prevalence instead of assigning these objects to a specific class. The lower bound includes only the confirmed and tentative fast rotators, while the upper bound also incorporates the unknown cases. In all cases, the denominator consists of the full set of observed targets, so that the resulting interval reflects the statistical content of the survey as fully as possible.

\begin{figure}[t!]
\centering
\includegraphics[width=\hsize]{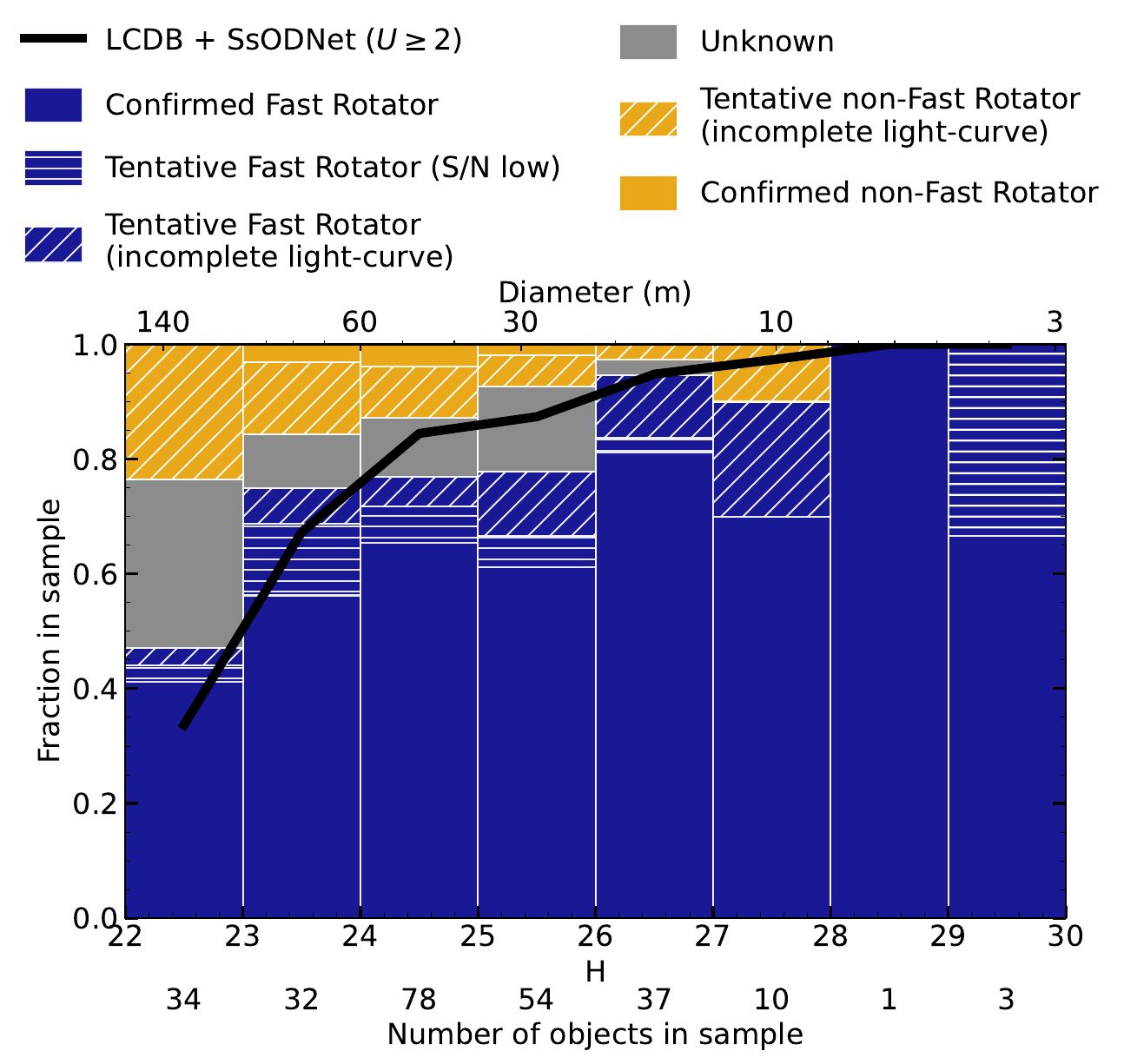}
  \caption{Histogram showing how the objects in our survey distribute by rotational classification as a function of absolute magnitude H. The blue bars indicate confirmed or tentative fast rotators, the orange bars show non-fast rotators, and the grey bars represent objects with insufficient information to determine their rotation state. The bottom horizontal axis displays the number of targets observed in each magnitude bin, whereas the top axis provides the corresponding diameter scale computed assuming an albedo of 0.15. The black line shows the fast-rotator fractions obtained from the LCDB and SsODNet ($U\geq2$), allowing a direct comparison with our sample.
}
     \label{fig:prev}
\end{figure}

Figure~\ref{fig:prev} shows how the sample distributes across the different determinations of rotational periods as a function of absolute magnitude. 
The fraction of objects classified as confirmed or tentative fast rotators increases steadily towards smaller sizes. At the same time, the proportion of unknown cases decreases towards faint magnitudes, mainly because the number of targets per magnitude bin becomes smaller in this regime.

The fast-rotator fractions for three absolute-magnitude ranges are summarised in Table~\ref{tab:prevalence}. 
For 22 < H < 24, the fast-rotator prevalence lies between $60.6\%$ and $80.3\%$, depending on how the unknown objects are treated. In the range 24 < H < 26, it increases to $77.3$--$89.4\%$, and for H > 26 it reaches $94.1$--$96.1\%$, indicating that nearly all objects in this smallest size domain are fast rotators. 
These values are consistent with those derived from the LCDB and SsODNet sample, indicating that our survey follows the same trend while extending it to smaller sizes, and constitutes, to our knowledge, the largest systematic campaign dedicated to measuring the rotation of such small NEAs.

\begin{table}
\centering
\caption{Prevalence of fast rotators in our sample compared with the LCDB and SsODNet.}
\label{tab:prevalence}
\begin{tabular}{ccccc}
\hline\hline
Magnitude & Total & FR & Prevalence & Ref. \\
\hline
22 < H < 24 & 66 & 32 (+8) & 60.6 -- 80.3 \% & 51.7 \% \\
24 < H < 26 & 132 & 84 (+18) & 77.3 -- 89.4 \% & 86.0 \% \\
H > 26 & 51 & 40 (+8) & 94.1 -- 96.1 \% & 96.6 \% \\
\hline
\end{tabular}
\tablefoot{
Fast rotators (FRs) include both confirmed plus tentative cases. Ref.: values compiled from LCDB and SsODNet ($U\ge 2$).
}
\end{table}

\section{Discussion}
\label{sec:discussion}

The distribution of rotation periods and sizes revealed by our survey bears directly on the internal structure and spin evolution of small NEAs. Our sample is not only rich in fast rotators, but also dominated by objects close to the detection limits of our facility, so any interpretation of the measured prevalence must begin by understanding the observational biases.

As illustrated in Fig.~\ref{fig:prev}, the fraction of objects classified as unknown is not uniform. For bright objects (H < 24), the higher fraction of unknowns is largely a result of the survey strategy. Given that non-fast rotators are statistically more common at larger sizes, our typical observing duration (40--120 minutes) is often insufficient to capture their full periodicity, thus preventing a clear classification. If an object has a rotation period significantly longer than the observing window, the resulting light curve may show a linear trend or no variation, thus preventing a secure period determination. This suggests that the unknown bin at bright magnitudes likely contains a higher proportion of slower rotators that were missed due to insufficient temporal coverage. In addition, most NEAs with absolute magnitude H between 22 and 23.5 were observed with the 0.80 m telescopes (Fig.~\ref{fig:ORs}), resulting in lower S/N, and therefore limiting our ability to identify them even as tentative non-fast rotators. As a consequence, the true fast-rotator fraction for the largest objects is expected to lie closer to the lower bound of our interval.

Towards smaller sizes, the character of the incompleteness changes. For H > 26, the observing runs usually span several expected rotation cycles, so temporal coverage is no longer the dominant limitation. Instead, the main sources of uncertainty are low signal-to-noise ratios and small light curve amplitudes. Spheroidal bodies with nearly circular cross-sections generate amplitudes comparable to the photometric scatter and become indistinguishable from noise. Asteroids with super-fast rotation can also exhibit artificially low amplitudes when the exposure time is a significant fraction of the rotation period, so that most of the intrinsic variability is smeared out and effectively averaged over each image. In this regime the survey is therefore incomplete primarily for low-amplitude fast-rotating objects, while high-amplitude fast rotators are efficiently recovered. Since the confirmed prevalence of fast rotators already exceeds $\sim$90 \% at these sizes, any correction for these biases would plausibly increase, rather than decrease, the inferred fraction. Taken together, the non-uniform distribution of unknowns with H implies that our lower limits of prevalence should be interpreted as conservative at both ends of the sampled size range.\\

The rotation periods of our sample span continuously from the classical cohesionless barrier at $\sim$2.2~h down to a few tens of seconds (Fig.~\ref{fig:spin_diagram}). Short periods become more frequent at small sizes, as expected in a strength-dominated regime, but fast rotators are spread across the entire domain without obvious gaps, clustering or trend. There is no indication of a preferred spin rate at which objects accumulate. Instead, the rotation periods appear to trace a continuous distribution from the gravity-defined limit down to the most extreme showing super-fast rotation. This homogeneity is interpreted as the outcome of torques that act over a broad range of timescales and do not favour a particular equilibrium. In such a scenario, individual NEAs wander through spin space under the combined action of YORP, tidal interactions, and impacts, with their current periods encoding a balance between these torques, internal strength, and the timing of the most recent disruptive events.

The spin rates we measure imply that a large fraction of the sample cannot be cohesionless rubble piles. Most objects lie well above the classical critical limit, so any physical interpretation of the period--diameter distribution must take the size uncertainties into account. Our diameters are derived from H assuming a geometric albedo $p_V = 0.15$, but the actual albedo distribution of small NEAs is far from uniform. Spectroscopic surveys of metre- to hectometre-sized NEAs show that S-complex objects dominate the taxonomic mix, typically accounting for $\sim$40--70\% of the population, with a significant contribution from X-complex and C/X-complex objects and a smaller fraction of other classes \citep{Devogele2019,Sanchez2024}. Within the X-complex, in particular, albedos span a wide range from dark primitive members (with $p_V \lesssim 0.1$) to very bright metal-rich asteroids (with $p_V$ commonly above $\sim 0.3$), and there is an observational bias towards discovering and characterising high-albedo objects at small sizes. If a substantial fraction of our sample were composed of dark objects with $p_V \sim 0.05$, their true sizes would be larger by a factor of 1.7, shifting the points to the right in Fig.~\ref{fig:spin_diagram} and moving more objects into the region that requires higher cohesive strength. Conversely, if high-albedo X- and S-complex asteroids with $p_V \sim 0.3-0.4$ dominated, the diameters would be smaller and the inferred structural demands correspondingly relaxed. Given the observed taxonomic distribution of small NEAs and the coexistence of both low- and high-albedo classes, our adopted albedo of 0.15 should be regarded as a statistical compromise rather than a precise value for individual objects. Within these bounds, the data are consistent with modest cohesive strengths of the order of tens to a few hundred pascals, comparable to those inferred for other small asteroids and for regolith on airless bodies, and the ensemble favours strength-dominated bodies with low but non-negligible cohesion over both perfectly strengthless aggregates and exceptionally strong monoliths.\\

Perhaps one of the most striking aspects of the sample is what is not observed. For H > 24, the overwhelming majority of objects with determined periods are fast rotators, and the fast-rotator fraction approaches unity for H > 26. 
In this small-size regime, our survey strategy would readily reveal slower rotators as strong monotonic trends or non-repeating segments within a given run, yet such cases are rare.

Our observing runs are not optimised to characterise slow rotators with periods of many hours, but such objects would still tend to appear as tentative low-frequency trends across a run rather than as high S/N, short-period signals. Among the smallest NEAs in our sample we find very few convincing cases of this kind, and the fraction of high S/N light curves that cannot be classified as fast or non-fast rotators is small. This suggests that genuinely slow rotators are intrinsically rare in the metre to decametre regime, and that small NEAs do not simply sample a scaled-down version of the spin distribution of kilometre-sized bodies. Instead, they appear to inhabit a regime in which YORP torques and recent reshaping events efficiently drive most objects into high-spin states. The lack of non-fast rotators at small sizes therefore appears to be a genuine physical property of the population, reinforcing the view that metre to decametre NEAs spend a substantial fraction of their lifetimes above the cohesionless spin limit.

Taken together, these results point to a scenario in which a large fraction of small NEAs are strength-dominated bodies with modest cohesion, repeatedly pushed towards and beyond the gravity-defined spin limit, and in which excited rotation is a frequent, perhaps recurrent, outcome rather than a transient anomaly. A more detailed analysis of the tumbling subsample, including quantitative constraints on damping timescales and excitation mechanisms, will be addressed in a forthcoming paper.\\

\begin{figure}[t!]
  \centering
  \begin{subfigure}{\columnwidth}
    \centering
    \includegraphics[width=\columnwidth]{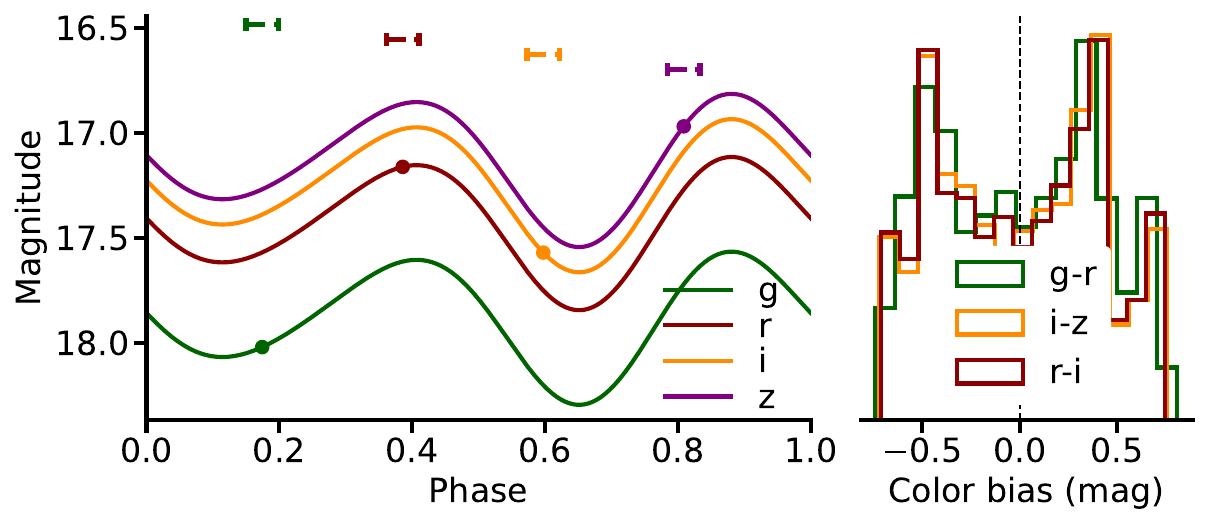}
    \caption{2025 HC: P$_{\rm rot} = 30.729 \pm 0.088$ s, T$_{\rm exp} = 1.5$ s.}
  \end{subfigure}
  \begin{subfigure}{\columnwidth}
    \centering
    \includegraphics[width=\columnwidth]{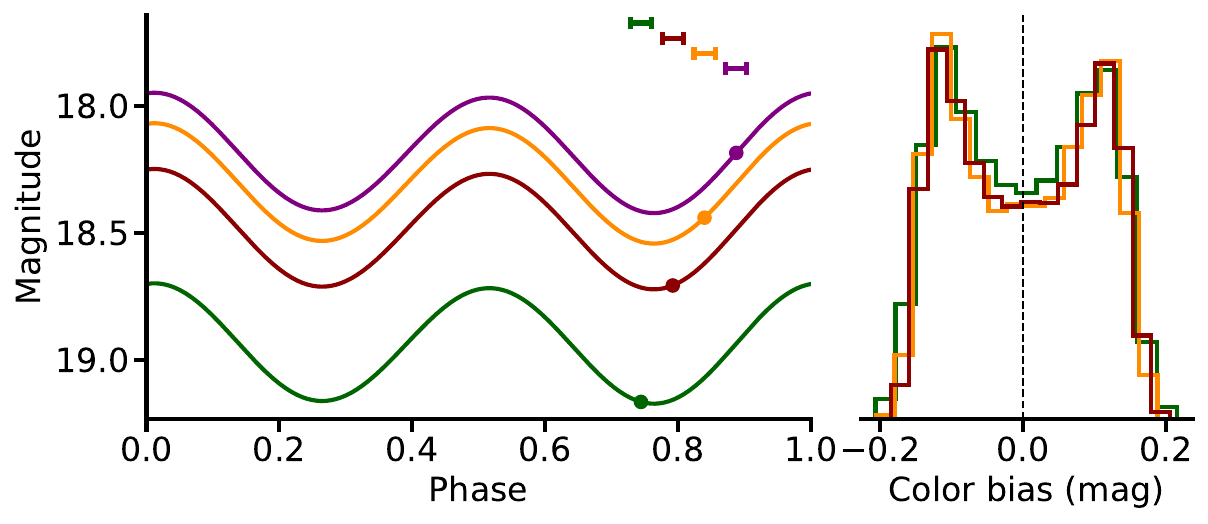}
    \caption{2025 OT4: P$_{\rm rot} = 5.228 \pm 0.081$ min, T$_{\rm exp} = 10$ s.}%
  \end{subfigure}
  \begin{subfigure}{\columnwidth}
    \centering
    \includegraphics[width=\columnwidth]{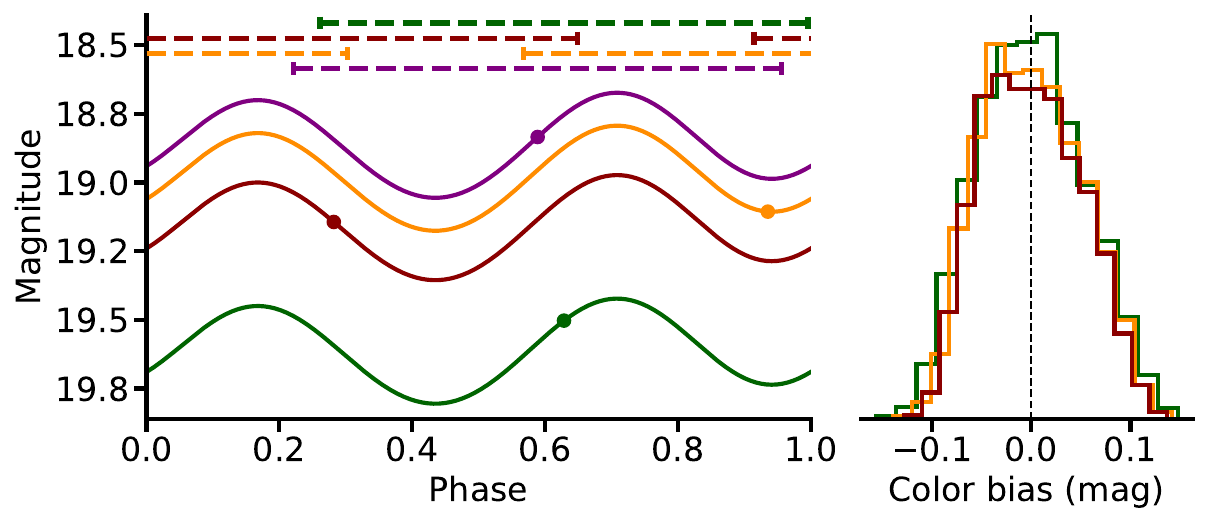}
    \caption{\label{fig:color_bias_c}2025 WR7: P$_{\rm rot} = 5.44 \pm 0.22$ s, T$_{\rm exp} = 4$ s.}%
  \end{subfigure}
\caption{Synthetic light curve models in the $g$, $r$, $i$, and $z$ bands are shown for three representative fast-rotating near-Earth asteroids. The left panel shows the per filter rotational models with an example of multi-filter observing cycle, supposing an inter-image overhead of 5 s; the exposure windows in rotational phase are indicated at the top of the panel by two vertical markers delimiting the start and end of each exposure, connected by a dashed line, while the circles mark the phase centres. The right panel shows the distribution of the colour bias $(\mathrm{mag}_X-\mathrm{mag}_Y)_{\mathrm{meas}} - (\mathrm{mag}_X-\mathrm{mag}_Y)_{\mathrm{true}}$ for 5000 random realisations of this sequence. The underlying true colours are assumed to be $(g-r) = 0.45$, $(r-i) = 0.18$, and $(i-z) = 0.12$; these values are illustrative and only set the absolute colour scale, without affecting the shape of the bias distributions. The histograms are plotted with arbitrary normalisation since only the shape of the distributions are relevant.
}
  \label{fig:color_bias}
\end{figure}

Finally, the prevalence of fast rotation at small sizes also has implications for other observational programmes that seek to characterise small NEAs. Surveys that derive colours, and hence taxonomic classifications, from non-simultaneous multi-filter photometry--as is the case, for instance, for LSST \citep{Jones2009,Kurlander2025}--will inevitably observe many fast rotators at different rotational phases in different filters. For objects with amplitudes of several tenths of a magnitude, which are quite common as we have shown here, sequential observations can easily sample distinct portions of the light curve, so that the measured colours reflect a combination of intrinsic spectral differences and rotational brightness variations. If uncorrected, this effect can inflate the apparent colour dispersion and bias objects towards or away from class boundaries. 

To illustrate this effect, we performed a simple experiment using three representative fast rotators drawn from our sample (see in Fig.~\ref{fig:color_bias}). For each object, we took the best-fitting phase-folded light curve model as a proxy for its rotational modulation and imposed an arbitrary but fixed colour offset between filters, corresponding to assumed intrinsic colours of $(g-r) = 0.45$, $(r-i) = 0.18$, and $(i-z) = 0.12$. We then simulated a non-simultaneous $g r i z$ observing sequence, repeated 5000 times with random starting phase, and computed the resulting colour pairs for each realisation. Comparing these measured colours to the imposed offsets yields a distribution of colour biases whose shape and width depend on the detailed light curve morphology and on the ratio of the rotation period to the effective exposure plus overhead time. For objects in the super-fast rotation regime, sequential multi-filter observations can therefore sample very different rotational phases in each band, naturally producing significant deviations from the true colours and, in turn, systematically degrading colour-based taxonomic classifications. In the particular case where the exposure time becomes comparable to, or a substantial fraction of, the rotation period, as in Fig.~\ref{fig:color_bias_c}, the bias in the derived colours can be significantly reduced, because the rotational flux variations are effectively averaged over the integration time.\\

Colour-based taxonomies of small NEAs are therefore intrinsically entangled with their spin properties, and truly unbiased taxonomic distributions will require  simultaneous multi-band observations, rotationally corrected colour measurements, or spectroscopic follow-up.

\section{Conclusions}\label{sec:conclusions}

We  conducted a dedicated survey of small near-Earth asteroids designed to characterise their rotational state using dense high-cadence light curves. To our knowledge, this is the first survey that targets the rotation status of such small NEAs in a systematic way, and it provides the largest homogeneous sample of fast rotators obtained by a single observational campaign to date. Our main results can be summarised as follows:

\begin{enumerate}
    \item We observed a total of 249 NEAs and derived reliable rotation periods for 161 of them, of which 156 exhibit $P < 2.2$~h and are therefore fast rotators. Among these objects, 91  show super-fast rotation, with periods shorter than 10~min, which places them among the most rapidly rotating NEAs known. When compared with the 520 fast rotators with $U \ge 2$ listed in the LCDB and SsoDNet, our survey increases the number of well-characterised fast-rotating NEAs by nearly one-third and provides a substantial expansion of the known sample in the sub-100~m regime.

    \item We quantified the prevalence of fast rotation as a function of absolute magnitude by combining confirmed, tentative, and unclassified objects into conservative lower and upper bounds. For 22 < H < 24, the fast-rotator fraction lies between 60.6\% and 80.3\%; in the range 24 < H < 26 it increases to 77.3--89.4\%, and for H > 26 it reaches 94.1--96.1\%, indicating that nearly all objects in this smallest size domain are fast rotators. These values are consistent with those derived from the LCDB and SsODNet sample, indicating that our survey follows the same trend while extending it to smaller sizes, and constitutes, to our knowledge, the largest systematic campaign dedicated to measuring the rotation of such small NEAs.

    \item The rotation periods in our sample span continuously from the classical spin barrier down to tens of seconds, with no evidence for a preferred spin rate or accumulation near specific periods. When compared with spin-limit curves derived for cohesive bodies, most objects lie above the gravity-defined critical limit and many require non-zero cohesion. In particular, 98 targets lie above the weak rubble-pile curve ($C \sim 10$~Pa), implying at least regolith-level cohesion or fractured monolithic structure, and 22 exceed the monolithic-strength curve and are only compatible with compact, high-strength interiors.

    \item Among the 156 fast rotators, we identified at least 35 objects displaying non-principal axis rotation (tumbling), nearly tripling the number of known fast-rotating tumblers when compared to the 12 such objects currently listed in the LCDB. This suggests that excited rotation is a common state among small NEAs and raises important questions about the true fraction of fast rotators that are in fact tumblers. A detailed analysis of this subsample, including the interplay between YORP-induced spin acceleration and the damping of non-principal axis rotation at very short periods, will be presented in a forthcoming paper.
\end{enumerate}

Overall, our results support a picture in which most small NEAs are strength-dominated bodies with modest but non-negligible cohesion, repeatedly driven towards and beyond the cohesionless spin limit. In this regime, fast and even excited rotation appear to be common outcomes rather than rare exceptions, with important consequences for the internal structure, collisional response, and surface evolution of metre- to decametre-sized near-Earth asteroids.

\section*{Data availability}

The derived rotational parameters and the calibrated photometric measurements for all objects presented in this work will be made available to the community through the LCDB.

\begin{acknowledgements}
This article is based on observations made in the Two-meter Twin Telescope (TTT) and the Transient Survey Telescope (TST) sited at the Teide Observatory of the Instituto de Astrofísica de Canarias (IAC), that Light Bridges operates in Tenerife, Canary Islands (Spain). The observation time rights (DTO) used for this research were consumed in the PEI "SIDERA2025". This research used storage and computing capacity in ASTRO POC's EDGE computing center at Tenerife under the form of Indefeasible Computer Rights (ICR), consumed in the PEI "SIDERA2025". Dr. Antonio Maudes’s insights in economics and law were instrumental in shaping the development of this work.
M.R.A. and J.L. acknowledge support from the Agencia Estatal de Investigaci\'on del Ministerio de Ciencia e Innovaci\'on (AEI-MCINN) under grant "Hydrated Minerals and Organic Compounds in Primitive Asteroids" with reference PID2020-120464GB-100. 
\end{acknowledgements}

\bibliographystyle{aa}
\bibliography{FR}

\begin{appendix}

\onecolumn

\clearpage
\section{Rotational parameters}

\setlength{\tabcolsep}{4pt}
\begin{longtable}{@{}lcccccccccc@{}}
\caption{\label{tab:periods}Rotational and observational parameters of the NEAs with confirmed principal axis rotation in our survey.}\\
\hline\hline
Object & P$_{\rm rot}$ & A (mag) & $a/b>$ & H (mag) & $\Delta$ (au) & $\alpha$ (deg) & T$_{\rm exp}$ (s) & Night & Tel. & Fig. (row, col)\\
\hline
\endfirsthead
\caption{Continued.}\\
\hline
Object & P$_{\rm rot}$ & A (mag) & $a/b>$ & H (mag) & $\Delta$ (au) & $\alpha$ (deg) & T$_{\rm exp}$ (s) & Night & Tel. & Fig. (row, col)\\
\hline
\endhead
\hline
\endfoot
2025 WR7 & $5.44 \pm 0.22$ s & 1.19 & 3.01 & 29.8 & 0.0038 & 34.7 & 4.0 & 2025-11-27 & TTT3 & \ref{fig:LC-1}, 1 (1,1)\\
2025 KS1 & $8.012 \pm 0.019$ s & 0.92 & 2.33 & 26.7 & 0.0144 & 44.0 & 1.5 & 2025-05-26 & TTT3 & \ref{fig:LC-1}, 1 (1,2)\\
2025 MP1 & $13.370 \pm 0.019$ s & 0.49 & 1.57 & 26.6 & 0.0116 & 60.7 & 3.5 & 2025-06-22 & TTT3 & \ref{fig:LC-1}, 1 (1,3)\\
2025 UF3 & $26.772 \pm 0.046$ s & 0.28 & 1.29 & 27.2 & 0.0166 & 18.7 & 5.7 & 2025-10-24 & TTT3 & \ref{fig:LC-1}, 1 (1,4)\\
2025 UK6 & $28.245 \pm 0.059$ s & 0.44 & 1.5 & 24.5 & 0.054 & 12.0 & 4.3 & 2025-10-29 & TTT3 & \ref{fig:LC-1}, 1 (1,5)\\
2025 HC & $30.729 \pm 0.088$ s & 0.73 & 1.96 & 26.4 & 0.0086 & 32.6 & 1.5 & 2025-04-19 & TTT3 & \ref{fig:LC-1}, 1 (2,1)\\
2025 SS5 & $31.833 \pm 0.022$ s & 0.19 & 1.19 & 27.8 & 0.0084 & 5.6 & 8.0 & 2025-09-26 & TTT1 & \ref{fig:LC-1}, 1 (2,2)\\
2025 FD2 & $32.31 \pm 0.79$ s & 0.23 & 1.24 & 27.5 & 0.0146 & 17.1 & 1.5 & 2025-03-25 & TTT3 & \ref{fig:LC-1}, 1 (2,3)\\
2025 OL1 & $37.816 \pm 0.041$ s & 0.27 & 1.28 & 25.0 & 0.0213 & 27.8 & 4.0 & 2025-07-26 & TTT1 & \ref{fig:LC-1}, 1 (2,4)\\
2025 MY1 & $44.406 \pm 0.099$ s & 0.36 & 1.39 & 23.1 & 0.0554 & 36.9 & 7.1 & 2025-06-27 & TTT2 & \ref{fig:LC-1}, 1 (2,5)\\
2025 FZ3 & $45.47 \pm 0.28$ s & 0.54 & 1.65 & 26.4 & 0.0216 & 36.6 & 3.0 & 2025-03-31 & TTT3 & \ref{fig:LC-1}, 1 (3,1)\\
2025 DL28 & $50.60 \pm 0.14$ s & 0.15 & 1.14 & 24.7 & 0.0476 & 37.3 & 8.0 & 2025-03-31 & TTT3 & \ref{fig:LC-1}, 1 (3,2)\\
2025 VW4 & $51.15 \pm 0.21$ s & 0.4 & 1.44 & 26.3 & 0.0246 & 21.7 & 6.0 & 2025-11-16 & TTT3 & \ref{fig:LC-1}, 1 (3,3)\\
2025 KC & $51.55 \pm 0.32$ s & 0.25 & 1.25 & 26.8 & 0.0182 & 44.3 & 6.7 & 2025-05-19 & TTT3 & \ref{fig:LC-1}, 1 (3,4)\\
2025 HV & $55.61 \pm 0.36$ s & 0.26 & 1.27 & 24.4 & 0.0539 & 38.7 & 4.0 & 2025-04-22 & TTT3 & \ref{fig:LC-1}, 1 (3,5)\\
2025 GL & $59.21 \pm 0.20$ s & 0.54 & 1.64 & 29.5 & 0.0098 & 12.4 & 6.0 & 2025-04-04 & TTT3 & \ref{fig:LC-1}, 1 (4,1)\\
2025 MJ & $1.0694 \pm 0.0069$ min & 0.23 & 1.23 & 24.9 & 0.0395 & 33.0 & 6.3 & 2025-06-21 & TTT3 & \ref{fig:LC-1}, 1 (4,2)\\
2025 VJ3 & $1.0749 \pm 0.0017$ min & 0.33 & 1.35 & 24.8 & 0.0233 & 4.3 & 3.0 & 2025-11-14 & TTT2 & \ref{fig:LC-1}, 1 (4,3)\\
2025 UT6 & $1.1415 \pm 0.0038$ min & 0.67 & 1.85 & 24.8 & 0.0497 & 28.5 & 19.4 & 2025-10-29 & TTT3 & \ref{fig:LC-1}, 1 (4,4)\\
2025 FT12 & $1.2261 \pm 0.0047$ min & 0.33 & 1.36 & 24.9 & 0.0353 & 19.2 & 5.0 & 2025-04-01 & TTT3 & \ref{fig:LC-1}, 1 (4,5)\\
2025 JW1 & $1.232 \pm 0.013$ min & 0.35 & 1.37 & 26.0 & 0.0181 & 23.4 & 3.0 & 2025-05-07 & TTT3 & \ref{fig:LC-1}, 1 (5,1)\\
2025 OM10 & $1.2898 \pm 0.0034$ min & 0.22 & 1.22 & 24.4 & 0.0227 & 20.1 & 6.1 & 2025-08-04 & TTT2 & \ref{fig:LC-1}, 1 (5,2)\\
2025 QO7 & $1.3410 \pm 0.0073$ min & 0.1 & 1.1 & 27.0 & 0.0074 & 21.4 & 4.1 & 2025-08-29 & TTT3 & \ref{fig:LC-1}, 1 (5,3)\\
2024 RB & $1.359 \pm 0.016$ min & 0.25 & 1.25 & 27.5 & 0.0128 & 4.8 & 5.0 & 2024-09-02 & TTT2 & \ref{fig:LC-1}, 1 (5,4)\\
2025 FL & $1.374 \pm 0.015$ min & 0.92 & 2.34 & 26.2 & 0.0075 & 103.8 & 3.0 & 2025-03-21 & TTT3 & \ref{fig:LC-1}, 1 (5,5)\\
2025 FW7 & $1.4386 \pm 0.0077$ min & 0.34 & 1.37 & 23.4 & 0.0669 & 50.0 & 4.0 & 2025-03-26 & TTT3 & \ref{fig:LC-1}, 1 (6,1)\\
2025 DT & $1.4659 \pm 0.0094$ min & 0.12 & 1.12 & 25.3 & 0.0418 & 3.7 & 6.0 & 2025-02-22 & TTT3 & \ref{fig:LC-1}, 1 (6,2)\\
2025 OM & $1.539 \pm 0.016$ min & 0.21 & 1.21 & 25.3 & 0.0445 & 16.6 & 14.0 & 2025-07-19 & TTT3 & \ref{fig:LC-1}, 1 (6,3)\\
2025 OW & $1.6169 \pm 0.0058$ min & 0.45 & 1.51 & 23.5 & 0.0463 & 45.0 & 19.0 & 2025-07-24 & TTT1 & \ref{fig:LC-1}, 1 (6,4)\\
2020 KT4 & $1.6512 \pm 0.0038$ min & 0.71 & 1.92 & 23.2 & 0.0769 & 9.8 & 10.1 & 2023-12-09 & TTT2 & \ref{fig:LC-1}, 1 (6,5)\\
2025 WH10 & $1.693 \pm 0.016$ min & 0.6 & 1.74 & 24.1 & 0.0529 & 27.8 & 10.1 & 2025-12-02 & TTT3 & \ref{fig:LC-1}, 1 (7,1)\\
2022 QD3 & $1.7213 \pm 0.0085$ min & 0.84 & 2.17 & 25.2 & 0.0415 & 15.3 & 6.2 & 2025-08-24 & TTT3 & \ref{fig:LC-1}, 1 (7,2)\\
2025 SF4 & $1.944 \pm 0.027$ min & 0.16 & 1.16 & 24.7 & 0.0441 & 6.6 & 5.4 & 2025-09-24 & TTT3 & \ref{fig:LC-1}, 1 (7,3)\\
2025 FQ4 & $1.959 \pm 0.012$ min & 0.46 & 1.52 & 22.7 & 0.058 & 54.1 & 10.0 & 2025-03-25 & TTT3 & \ref{fig:LC-1}, 1 (7,4)\\
2017 MB3 & $1.9846 \pm 0.0054$ min & 1.24 & 3.14 & 25.4 & 0.0129 & 85.3 & 10.0 & 2024-06-30 & TTT2 & \ref{fig:LC-1}, 1 (7,5)\\
2024 OM1 & $2.0342 \pm 0.0057$ min & 0.28 & 1.3 & 24.2 & 0.0292 & 18.9 & 10.0 & 2024-08-02 & TTT2 & \ref{fig:LC-1}, 1 (8,1)\\
2025 WA & $2.043 \pm 0.011$ min & 0.58 & 1.71 & 24.9 & 0.0262 & 31.3 & 4.2 & 2025-11-17 & TTT3 & \ref{fig:LC-1}, 1 (8,2)\\
2023 NT1 & $2.1182 \pm 0.0034$ min & 0.68 & 1.86 & 25.0 & 0.0224 & 29.4 & 18.3 & 2023-07-16 & TTT2 & \ref{fig:LC-1}, 1 (8,3)\\
2025 OE & $2.212 \pm 0.031$ min & 0.34 & 1.36 & 24.1 & 0.0493 & 31.9 & 7.0 & 2025-07-18 & TTT3 & \ref{fig:LC-1}, 1 (8,4)\\
2025 FN6 & $2.252 \pm 0.031$ min & 0.6 & 1.73 & 23.9 & 0.0817 & 16.3 & 8.0 & 2025-04-22 & TTT3 & \ref{fig:LC-1}, 1 (8,5)\\
2024 CB1 & $2.692 \pm 0.016$ min & 0.22 & 1.22 & 24.2 & 0.0263 & 23.4 & 1.0 & 2024-02-10 & TTT2 & \ref{fig:LC-1}, 1 (9,1)\\
2019 KK5 & $2.768 \pm 0.045$ min & 0.57 & 1.69 & 22.8 & 0.0794 & 25.5 & 2.2 & 2024-01-08 & TTT2 & \ref{fig:LC-1}, 1 (9,2)\\
2025 MV88 & $2.872 \pm 0.042$ min & 0.24 & 1.25 & 25.0 & 0.0281 & 20.0 & 1.6 & 2025-06-27 & TTT3 & \ref{fig:LC-1}, 1 (9,3)\\
2025 JR & $2.979 \pm 0.026$ min & 0.33 & 1.35 & 23.2 & 0.0746 & 22.2 & 15.3 & 2025-05-17 & TTT3 & \ref{fig:LC-1}, 1 (9,4)\\
2024 CG2 & $3.251 \pm 0.026$ min & 1.14 & 2.86 & 23.1 & 0.0268 & 37.5 & 6.9 & 2024-02-12 & TTT2 & \ref{fig:LC-1}, 1 (9,5)\\
2025 BB2 & $3.319 \pm 0.027$ min & 0.54 & 1.65 & 25.6 & 0.0161 & 60.8 & 25.0 & 2025-01-30 & TTT2 & \ref{fig:LC-2}, 2 (1,1)\\
2025 FS5 & $3.531 \pm 0.043$ min & 0.13 & 1.13 & 24.2 & 0.0505 & 35.5 & 2.5 & 2025-03-25 & TTT3 & \ref{fig:LC-2}, 2 (1,2)\\
2025 LV & $3.605 \pm 0.047$ min & 0.27 & 1.29 & 24.8 & 0.0414 & 13.0 & 8.5 & 2025-06-17 & TTT3 & \ref{fig:LC-2}, 2 (1,3)\\
2025 FS & $3.793 \pm 0.060$ min & 0.57 & 1.69 & 26.0 & 0.0198 & 22.0 & 3.0 & 2025-03-21 & TTT3 & \ref{fig:LC-2}, 2 (1,4)\\
2025 GQ1 & $3.919 \pm 0.047$ min & 0.48 & 1.56 & 25.8 & 0.0269 & 18.1 & 10.0 & 2025-04-19 & TTT3 & \ref{fig:LC-2}, 2 (1,5)\\
2025 DS2 & $4.037 \pm 0.038$ min & 0.35 & 1.38 & 24.5 & 0.0297 & 9.9 & 4.0 & 2025-02-25 & TTT3 & \ref{fig:LC-2}, 2 (2,1)\\
2025 HB1 & $4.470 \pm 0.052$ min & 0.25 & 1.26 & 24.2 & 0.0457 & 38.4 & 4.0 & 2025-04-23 & TTT3 & \ref{fig:LC-2}, 2 (2,2)\\
2025 DU25 & $4.627 \pm 0.020$ min & 0.71 & 1.93 & 27.2 & 0.0077 & 39.7 & 1.5 & 2025-03-17 & TTT3 & \ref{fig:LC-2}, 2 (2,3)\\
2024 QS1 & $4.803 \pm 0.034$ min & 0.67 & 1.85 & 22.0 & 0.0817 & 41.2 & 10.0 & 2024-09-02 & TTT2 & \ref{fig:LC-2}, 2 (2,4)\\
2025 TU1 & $4.814 \pm 0.070$ min & 0.3 & 1.32 & 26.5 & 0.0186 & 13.1 & 2.7 & 2025-10-10 & TTT3 & \ref{fig:LC-2}, 2 (2,5)\\
2025 SZ4 & $5.058 \pm 0.025$ min & 0.75 & 1.99 & 23.9 & 0.0249 & 31.0 & 3.0 & 2025-09-25 & TTT1 & \ref{fig:LC-2}, 2 (3,1)\\
2025 OT4 & $5.228 \pm 0.081$ min & 0.47 & 1.55 & 24.8 & 0.0256 & 42.2 & 10.0 & 2025-07-31 & TTT3 & \ref{fig:LC-2}, 2 (3,2)\\
2025 WH20 & $5.380 \pm 0.083$ min & 0.18 & 1.18 & 25.7 & 0.0291 & 5.4 & 3.2 & 2025-12-10 & TTT3 & \ref{fig:LC-2}, 2 (3,3)\\
2023 RF13 & $5.473 \pm 0.088$ min & 0.55 & 1.65 & 25.3 & 0.0322 & 24.8 & 3.3 & 2023-09-23 & TTT1 & \ref{fig:LC-2}, 2 (3,4)\\
2015 XR1 & $5.600 \pm 0.073$ min & 0.2 & 1.2 & 23.2 & 0.144 & 17.9 & 28.7 & 2025-05-24 & TTT3 & \ref{fig:LC-2}, 2 (3,5)\\
2025 RX4 & $6.019 \pm 0.036$ min & 0.46 & 1.53 & 24.7 & 0.0216 & 48.3 & 5.3 & 2025-09-21 & TTT3 & \ref{fig:LC-2}, 2 (4,1)\\
2023 SN5 & $6.32 \pm 0.13$ min & 0.27 & 1.29 & 25.0 & 0.0243 & 19.8 & 4.2 & 2023-09-25 & TTT1 & \ref{fig:LC-2}, 2 (4,2)\\
2025 UD & $6.333 \pm 0.037$ min & 0.6 & 1.74 & 24.8 & 0.019 & 39.9 & 1.4 & 2025-10-18 & TTT3 & \ref{fig:LC-2}, 2 (4,3)\\
2025 KE & $6.657 \pm 0.095$ min & 0.2 & 1.2 & 23.9 & 0.0313 & 18.8 & 4.0 & 2025-05-20 & TTT3 & \ref{fig:LC-2}, 2 (4,4)\\
2025 OA3 & $6.767 \pm 0.095$ min & 0.79 & 2.06 & 24.9 & 0.0172 & 57.1 & 1.3 & 2025-08-02 & TTT3 & \ref{fig:LC-2}, 2 (4,5)\\
2022 MA3 & $6.78 \pm 0.11$ min & 0.59 & 1.73 & 22.2 & 0.0567 & 52.3 & 5.0 & 2025-07-08 & TTT2 & \ref{fig:LC-2}, 2 (5,1)\\
2025 DE1 & $7.225 \pm 0.039$ min & 0.19 & 1.19 & 25.0 & 0.0233 & 18.2 & 10.0 & 2025-02-22 & TTT3 & \ref{fig:LC-2}, 2 (5,2)\\
2025 SA73 & $8.027 \pm 0.020$ min & 0.32 & 1.35 & 26.6 & 0.0204 & 15.3 & 5.1 & 2025-10-12 & TTT3 & \ref{fig:LC-2}, 2 (5,3)\\
2025 KB & $8.04 \pm 0.16$ min & 0.19 & 1.19 & 24.6 & 0.041 & 47.8 & 7.8 & 2025-05-17 & TTT3 & \ref{fig:LC-2}, 2 (5,4)\\
2025 FV & $8.08 \pm 0.12$ min & 0.33 & 1.35 & 22.3 & 0.1624 & 0.5 & 34.0 & 2025-05-01 & TTT2 & \ref{fig:LC-2}, 2 (5,5)\\
2023 UO & $8.213 \pm 0.095$ min & 0.29 & 1.31 & 25.0 & 0.026 & 22.9 & 1.6 & 2023-11-09 & TTT2 & \ref{fig:LC-2}, 2 (6,1)\\
2025 FN14 & $9.103 \pm 0.058$ min & 0.36 & 1.4 & 24.9 & 0.0503 & 35.5 & 8.0 & 2025-03-31 & TTT3 & \ref{fig:LC-2}, 2 (6,2)\\
2024 PC2 & $10.83 \pm 0.18$ min & 0.33 & 1.35 & 22.2 & 0.0486 & 48.8 & 7.6 & 2024-08-08 & TTT2 & \ref{fig:LC-2}, 2 (6,3)\\
2025 KK2 & $10.90 \pm 0.17$ min & 0.32 & 1.35 & 24.5 & 0.0882 & 10.5 & 15.8 & 2025-05-25 & TTT3 & \ref{fig:LC-2}, 2 (6,4)\\
2025 HQ1 & $12.03 \pm 0.36$ min & 0.29 & 1.3 & 25.8 & 0.0335 & 34.6 & 10.0 & 2025-04-25 & TTT3 & \ref{fig:LC-2}, 2 (6,5)\\
2023 NO1 & $12.27 \pm 0.25$ min & 0.48 & 1.55 & 22.7 & 0.1179 & 29.0 & 12.9 & 2023-08-13 & TTT2 & \ref{fig:LC-2}, 2 (7,1)\\
2025 TV10 & $12.56 \pm 0.17$ min & 0.82 & 2.13 & 25.1 & 0.0206 & 52.5 & 2.6 & 2025-10-18 & TTT3 & \ref{fig:LC-2}, 2 (7,2)\\
2023 OJ1 & $12.8300 \pm 0.0077$ min & 0.88 & 2.25 & 26.1 & 0.0245 & 38.9 & 23.9 & 2023-07-24 & TTT1 & \ref{fig:LC-2}, 2 (7,3)\\
2011 YU74 & $14.57 \pm 0.27$ min & 2.19 & 7.5 & 22.8 & 0.1501 & 2.7 & 40.5 & 2025-06-27 & TTT2 & \ref{fig:LC-2}, 2 (7,4)\\
2025 DA15 & $17.70 \pm 0.21$ min & 0.32 & 1.34 & 24.9 & 0.048 & 25.7 & 6.0 & 2025-03-18 & TTT3 & \ref{fig:LC-2}, 2 (7,5)\\
2025 KZ & $18.02 \pm 0.26$ min & 0.38 & 1.42 & 26.3 & 0.0251 & 3.8 & 6.3 & 2025-05-21 & TTT3 & \ref{fig:LC-2}, 2 (8,1)\\
2009 HC & $18.614 \pm 0.088$ min & 0.18 & 1.18 & 24.6 & 0.0439 & 36.9 & 21.2 & 2025-10-10 & TTT3 & \ref{fig:LC-2}, 2 (8,2)\\
2025 KV4 & $18.69 \pm 0.34$ min & 0.45 & 1.52 & 25.6 & 0.0284 & 53.6 & 10.5 & 2025-06-06 & TTT3 & \ref{fig:LC-2}, 2 (8,3)\\
2025 RH2 & $18.732 \pm 0.036$ min & 0.23 & 1.23 & 24.9 & 0.0608 & 10.2 & 14.4 & 2025-09-22 & TTT3 & \ref{fig:LC-2}, 2 (8,4)\\
2025 VU & $18.896 \pm 0.092$ min & 0.46 & 1.53 & 26.5 & 0.02 & 6.8 & 3.3 & 2025-11-10 & TTT3 & \ref{fig:LC-2}, 2 (8,5)\\
2025 ES1 & $20.04 \pm 0.33$ min & 1.51 & 4.01 & 24.0 & 0.0352 & 53.4 & 5.0 & 2025-03-09 & TTT3 & \ref{fig:LC-2}, 2 (9,1)\\
2025 SD4 & $20.79 \pm 0.15$ min & 0.42 & 1.48 & 25.9 & 0.0194 & 24.8 & 1.6 & 2025-09-24 & TTT3 & \ref{fig:LC-2}, 2 (9,2)\\
2025 DR4 & $21.23 \pm 0.36$ min & 0.23 & 1.24 & 26.2 & 0.0309 & 11.8 & 10.0 & 2025-02-25 & TTT3 & \ref{fig:LC-2}, 2 (9,3)\\
2025 VZ1 & $21.27 \pm 0.18$ min & 0.51 & 1.61 & 24.3 & 0.0459 & 56.8 & 13.3 & 2025-11-16 & TTT3 & \ref{fig:LC-2}, 2 (9,4)\\
2025 RK & $21.57 \pm 0.11$ min & 0.34 & 1.37 & 24.8 & 0.0322 & 16.2 & 3.2 & 2025-09-03 & TTT3 & \ref{fig:LC-2}, 2 (9,5)\\
2025 WG & $22.04 \pm 0.25$ min & 0.13 & 1.13 & 26.5 & 0.0212 & 5.5 & 4.6 & 2025-11-27 & TTT3 & \ref{fig:LC-3}, 3 (1,1)\\
2025 DM3 & $22.22 \pm 0.10$ min & 0.75 & 2.0 & 24.2 & 0.0527 & 24.3 & 6.0 & 2025-03-09 & TTT3 & \ref{fig:LC-3}, 3 (1,2)\\
2025 HZ6 & $22.42 \pm 0.34$ min & 1.63 & 4.5 & 24.7 & 0.0773 & 8.5 & 12.0 & 2025-04-30 & TTT3 & \ref{fig:LC-3}, 3 (1,3)\\
2025 QB5 & $26.453 \pm 0.092$ min & 0.36 & 1.4 & 25.7 & 0.0347 & 18.3 & 8.5 & 2025-09-01 & TTT3 & \ref{fig:LC-3}, 3 (1,4)\\
2025 PR1 & $26.82 \pm 0.44$ min & 0.24 & 1.25 & 27.1 & 0.0115 & 3.6 & 2.0 & 2025-08-14 & TTT1 & \ref{fig:LC-3}, 3 (1,5)\\
2025 XV & $27.11 \pm 0.50$ min & 0.42 & 1.48 & 22.8 & 0.0441 & 39.1 & 10.0 & 2025-12-10 & TTT2 & \ref{fig:LC-3}, 3 (2,1)\\
2025 SW & $27.782 \pm 0.057$ min & 1.73 & 4.9 & 22.9 & 0.0208 & 65.8 & 5.0 & 2025-09-21 & TTT1 & \ref{fig:LC-3}, 3 (2,2)\\
2025 CB1 & $27.79 \pm 0.35$ min & 0.9 & 2.29 & 24.5 & 0.0938 & 20.2 & 60.0 & 2025-01-26 & TST & \ref{fig:LC-3}, 3 (2,3)\\
2025 SX2 & $28.84 \pm 0.14$ min & 0.28 & 1.29 & 26.1 & 0.0132 & 34.7 & 3.1 & 2025-09-21 & TTT3 & \ref{fig:LC-3}, 3 (2,4)\\
2025 FU5 & $29.13 \pm 0.41$ min & 1.81 & 5.3 & 23.2 & 0.0544 & 46.1 & 12.0 & 2025-05-18 & TTT2 & \ref{fig:LC-3}, 3 (2,5)\\
2025 WT & $31.46 \pm 0.22$ min & 0.15 & 1.15 & 24.7 & 0.0407 & 31.1 & 3.8 & 2025-11-28 & TTT3 & \ref{fig:LC-3}, 3 (3,1)\\
2025 OK1 & $34.72 \pm 0.45$ min & 0.19 & 1.19 & 25.3 & 0.0148 & 25.7 & 2.0 & 2025-07-23 & TTT1 & \ref{fig:LC-3}, 3 (3,2)\\
2025 TQ10 & $35.40 \pm 0.14$ min & 1.64 & 4.51 & 24.5 & 0.024 & 61.5 & 3.0 & 2025-10-20 & TTT3 & \ref{fig:LC-3}, 3 (3,3)\\
2025 DO & $36.1 \pm 1.2$ min & 0.87 & 2.24 & 25.5 & 0.017 & 44.1 & 5.0 & 2025-02-20 & TTT3 & \ref{fig:LC-3}, 3 (3,4)\\
2025 EA1 & $38.69 \pm 0.54$ min & 0.42 & 1.47 & 24.9 & 0.0513 & 6.0 & 5.0 & 2025-03-09 & TTT3 & \ref{fig:LC-3}, 3 (3,5)\\
2025 VS & $40.69 \pm 0.39$ min & 0.16 & 1.16 & 26.3 & 0.0197 & 7.3 & 3.7 & 2025-11-10 & TTT3 & \ref{fig:LC-3}, 3 (4,1)\\
2025 KU6 & $42.19 \pm 0.92$ min & 0.5 & 1.58 & 23.9 & 0.0647 & 7.5 & 7.7 & 2025-06-13 & TTT3 & \ref{fig:LC-3}, 3 (4,2)\\
2024 RJ32 & $47.06 \pm 0.63$ min & 0.29 & 1.31 & 25.0 & 0.0202 & 28.3 & 4.0 & 2024-10-03 & TST & \ref{fig:LC-3}, 3 (4,3)\\
2024 RE & $47.24 \pm 0.52$ min & 0.22 & 1.22 & 23.4 & 0.0462 & 72.3 & 10.0 & 2024-09-03 & TTT2 & \ref{fig:LC-3}, 3 (4,4)\\
2025 JE1 & $48.7 \pm 1.1$ min & 0.47 & 1.53 & 25.8 & 0.0323 & 5.9 & 25.0 & 2025-05-06 & TTT2 & \ref{fig:LC-3}, 3 (4,5)\\
2020 XT2 & $48.93 \pm 0.89$ min & 0.42 & 1.47 & 24.8 & 0.0629 & 34.4 & 16.0 & 2025-04-18 & TTT3 & \ref{fig:LC-3}, 3 (5,1)\\
2025 OL & $54.38 \pm 0.77$ min & 0.23 & 1.23 & 25.1 & 0.0169 & 34.5 & 6.0 & 2025-07-19 & TTT2 & \ref{fig:LC-3}, 3 (5,2)\\
2023 SU & $65.14 \pm 0.30$ min & 0.67 & 1.86 & 24.2 & 0.0614 & 28.8 & 13.7 & 2023-10-16 & TTT2 & \ref{fig:LC-3}, 3 (5,3)\\
2025 TK11 & $70.71 \pm 0.97$ min & 0.4 & 1.45 & 24.2 & 0.1002 & 13.7 & 30.7 & 2025-11-02 & TTT3 & \ref{fig:LC-3}, 3 (5,4)\\
2024 XA8 & $85.6 \pm 1.1$ min & 0.29 & 1.31 & 22.9 & 0.1177 & 19.9 & 10.0 & 2025-03-28 & TTT3 & \ref{fig:LC-3}, 3 (5,5)\\
2025 KY4 & $90.2 \pm 1.9$ min & 0.33 & 1.35 & 26.8 & 0.0231 & 10.8 & 7.0 & 2025-05-29 & TTT3 & \ref{fig:LC-3}, 3 (6,1)\\
2024 TS8 & $96.89 \pm 0.39$ min & 1.12 & 2.81 & 23.0 & 0.0826 & 23.5 & 20.0 & 2024-12-19 & TST & \ref{fig:LC-3}, 3 (6,2)\\
2025 PE & $100.2 \pm 2.2$ min & 1.04 & 2.6 & 23.9 & 0.042 & 32.3 & 23.0 & 2025-08-04 & TTT2 & \ref{fig:LC-3}, 3 (6,3)\\
2025 OO & $100.4 \pm 2.2$ min & 0.62 & 1.77 & 24.2 & 0.06 & 3.8 & 9.2 & 2025-07-20 & TTT3 & \ref{fig:LC-3}, 3 (6,4)\\
2025 PM & $114.5 \pm 1.6$ min & 0.48 & 1.56 & 24.2 & 0.1058 & 7.0 & 30.0 & 2025-08-03 & TTT3 & \ref{fig:LC-3}, 3 (6,5)\\
2023 RW9 & $126.6 \pm 6.1$ min & 0.33 & 1.35 & 22.0 & 0.1316 & 22.6 & 5.3 & 2023-09-18 & TTT1 & \ref{fig:LC-3}, 3 (7,1)\\
2025 PX2 & $132.77 \pm 0.38$ min & 0.45 & 1.51 & 24.5 & 0.0896 & 8.0 & 14.7 & 2025-09-11 & TTT3 & \ref{fig:LC-3}, 3 (7,2)\\
2025 QP3 & $133.14 \pm 0.34$ min & 0.9 & 2.28 & 24.5 & 0.0182 & 24.9 & 7.4 & 2025-08-24 & TTT3 & \ref{fig:LC-3}, 3 (7,3)\\
2016 QF84 & $137.5 \pm 1.1$ min & 0.78 & 2.05 & 23.3 & 0.0748 & 9.5 & 11.0 & 2025-08-20 & TTT1 & \ref{fig:LC-3}, 3 (7,4)\\
2023 RF1 & $148.71 \pm 0.71$ min & 0.72 & 1.95 & 25.9 & 0.0496 & 7.9 & 18.2 & 2023-09-17 & TTT2 & \ref{fig:LC-3}, 3 (7,5)\\
2020 UQ3 & $197.8 \pm 7.5$ min & 0.83 & 2.15 & 24.1 & 0.0474 & 42.1 & 3.3 & 2023-07-09 & TTT1 & \ref{fig:LC-3}, 3 (8,1)\\
\end{longtable}
\tablefoot{
P$_{\rm rot}$ is the rotation period derived in this work; A is the peak-to-peak light curve amplitude, corrected for exposure time smearing effects. The axis ratio lower limit $a/b$ is computed from A assuming triaxial ellipsoids viewed equator-on. H is the absolute magnitude from the JPL Horizons system at the time of observation. $\Delta$ and $\alpha$ denote the geocentric distance and solar phase angle at mid-exposure, respectively. T$_{\rm exp}$ is the exposure time used for individual frames. Night gives the UT date of the beginning of the night, and Tel. indicates the telescope used. The last column specifies the panel in Fig.~\ref{fig:LC-1} where the corresponding phase-folded light curve is shown. Tumblers and objects without a secure period are not listed in this table.
}
\setlength{\tabcolsep}{6pt}

\FloatBarrier

\section{\label{sec:append_LC}Phased-folded light curves}

\begin{figure}
\centering
\includegraphics[width=\textwidth]{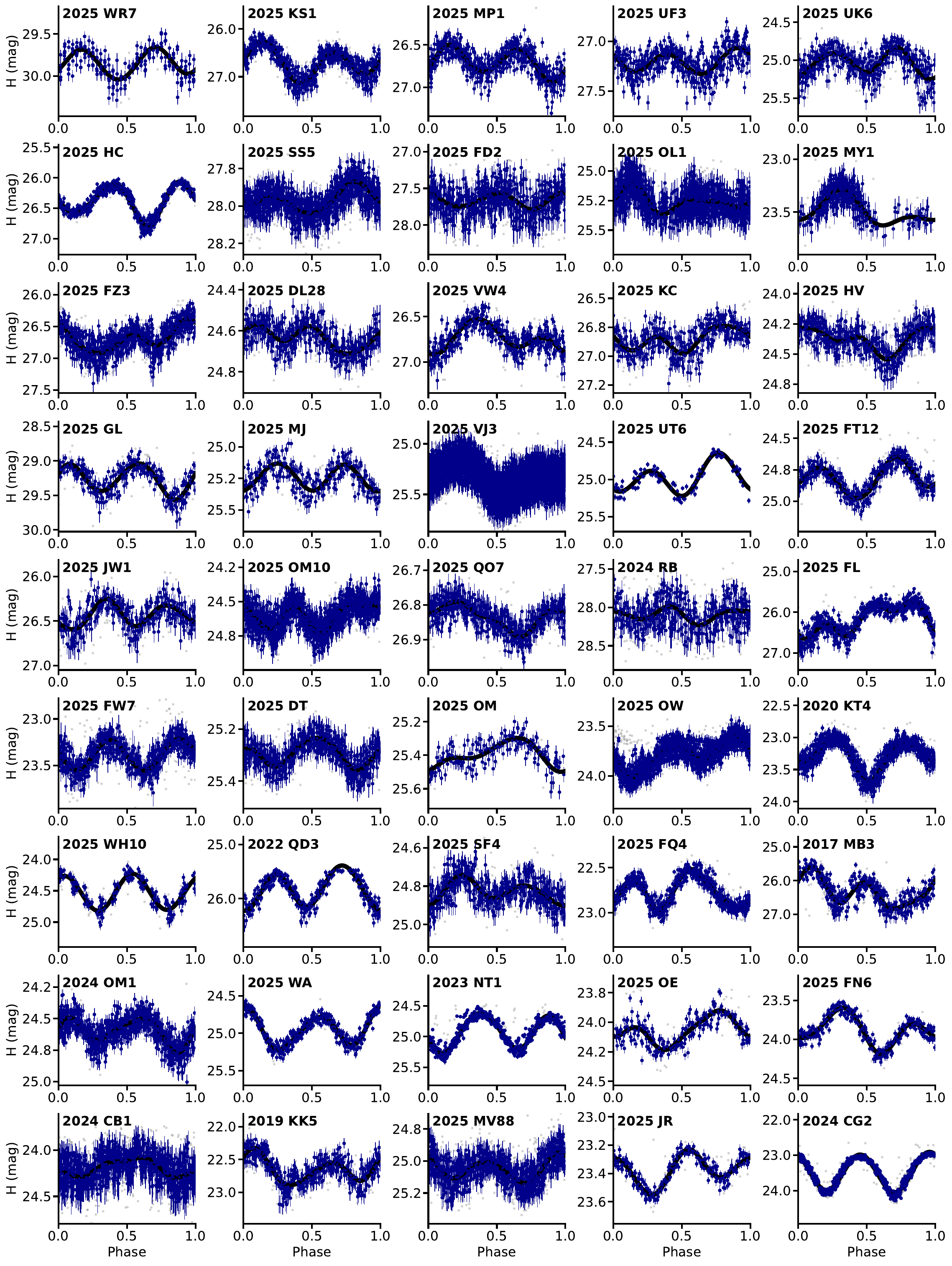}
\caption{Phase-folded light curves of the confirmed principal axis rotators.}
\label{fig:LC-1}
\end{figure}

\begin{figure}
\ContinuedFloat
\centering
\includegraphics[width=\textwidth]{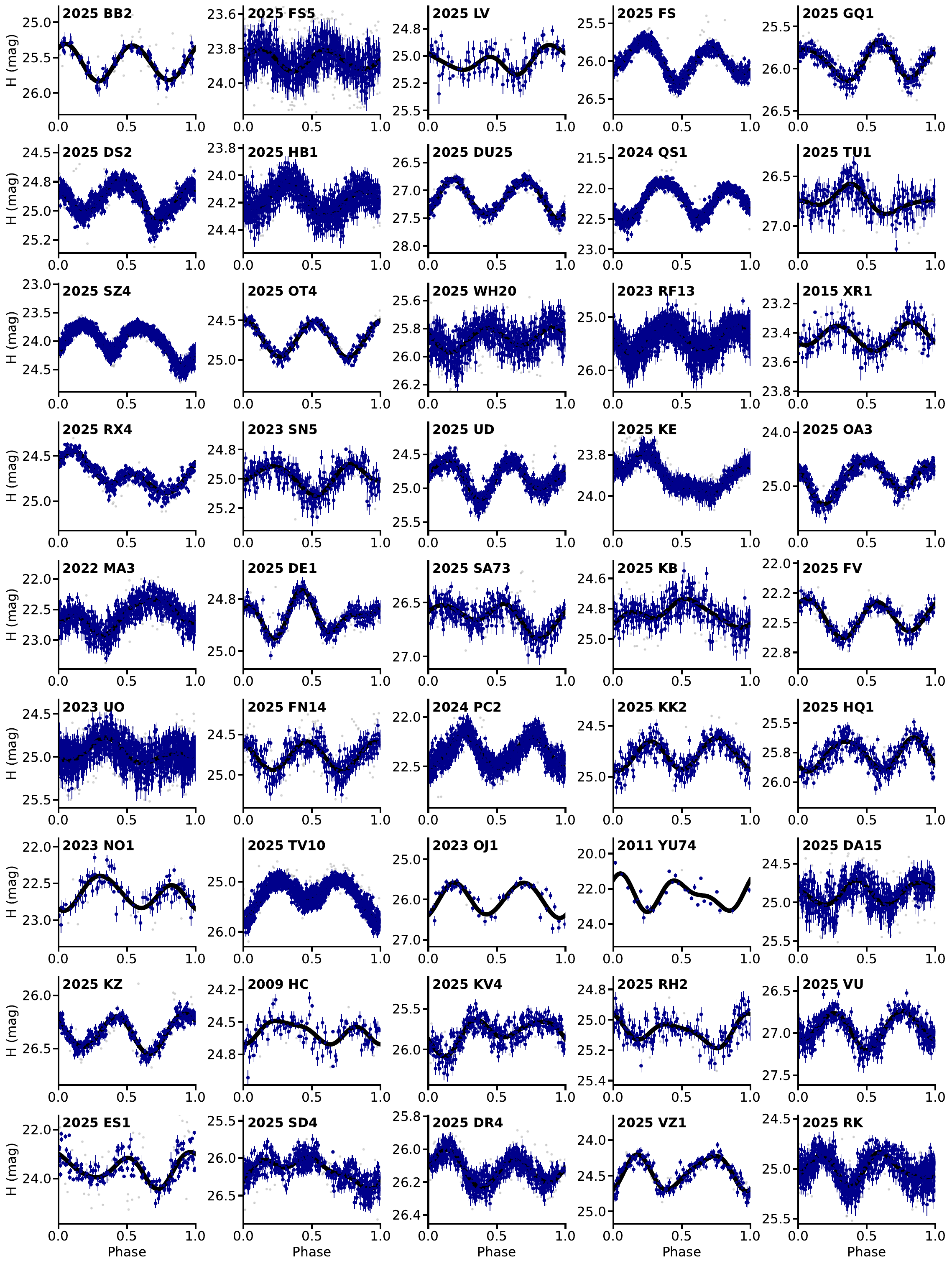}
\caption{Continued.}
\label{fig:LC-2}
\end{figure}

\begin{figure}
\ContinuedFloat
\centering
\includegraphics[width=\textwidth]{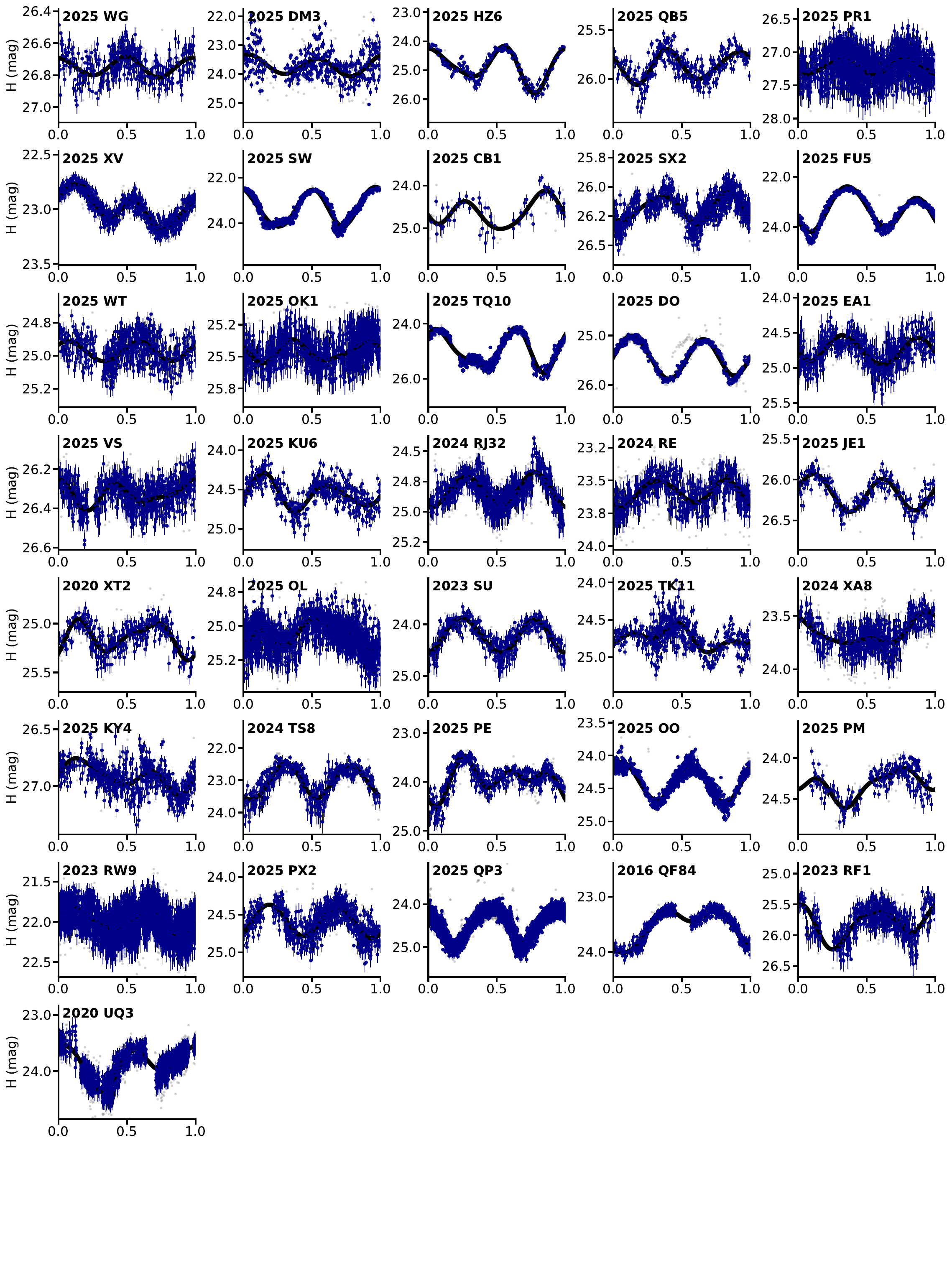}
\caption{Continued.}
\label{fig:LC-3}
\end{figure}

\FloatBarrier

\end{appendix}
\end{document}